%% file: documento.tex
\documentclass[        
    a4paper,          
    12pt,             
    chapter=TITLE,    
    section=Title,    
    subsection=Title, 
    oneside,          
    english,          
    spanish,          
    brazil,           
    fleqn             
]{abntex2}

\input{lib/preambulo}

\trabalhoacademico{tccgraduacao}

\ehqualificacao{nao}

\removerbordasdohyperlink{sim} 

\cordohyperlink{nao}

\ies{Instituto Federal de Educação, Ciência e Tecnologia do Ceará}
\iessigla{IFCE}
\centro{Campus Acaraú}

\graduacaoem{Física}
\habilitacao{licenciado}

\especializacaoem{Yyyyyyyyy}

\programamestrado{Programa de Pós-Graduação em Xxxxxxx}
\nomedomestrado{Mestrado Acadêmico em Xxxxxxx}
\mestreem{Engenharia Xxxxxx}
\areadeconcentracaomestrado{Engenharia Xxxxxx}

\programadoutorado{Programa de Pós-Graduação em Xxxxxx}
\nomedodoutorado{Doutorado em Xxxxxxx}
\doutorem{Engenharia Xxxxxx}
\areadeconcentracaodoutorado{Engenharia Xxxxxxx}

\autor{Marcos Vinícius dos Santos Araújo}
\titulo{Aplicação do Movimento Browniano Geométrico para simulação de preços de ações do índice brasileiro de \emph{Small Caps}}
\data{2020}
\local{Acaraú}

\dataaprovacao{09/10/2020}

\orientador{Dr. César Menezes Vieira}
\orientadories{Instituto Federal de Educação, Ciência e Tecnologia do Ceará (IFCE)}
\orientadorfeminino{nao} 

\coorientador{Dr. Paulo Henrique Nobre Parente}
\coorientadories{Instituto Federal de Educação, Ciência e Tecnologia do Ceará (IFCE)}
\coorientadorfeminino{nao} 

\membrodabancadois{Dr. Hygor Piaget Monteiro Melo}
\membrodabancadoiscentro{Centro de Física Teórica e Computacional}
\membrodabancadoisies{Universidade de Lisboa (ULisboa)}

\begin{document}	

	\imprimircapa
	\imprimirfolhaderosto{}
	\imprimirfichacatalografica{1-pre-textuais/ficha-catalografica}
	\imprimirfolhadeaprovacao
	\input{1-pre-textuais/dedicatoria}
	\input{1-pre-textuais/agradecimentos}
	\input{1-pre-textuais/epigrafe}
	 %
	\begin{resumo} %
		\input{1-pre-textuais/resumo} %
	\end{resumo} %

	\begin{resumo}[Abstract] %
		\input{1-pre-textuais/abstract} %
	\end{resumo} %

	\renewcommand*\listfigurename{Lista de Figuras} 
	\imprimirlistadeilustracoes
	\imprimirlistadetabelas
	\imprimirsumario
	
	\setcounter{table}{0}
	
	\textual
	\input{2-textuais/1-introducao}

	\input{2-textuais/2-fundamentacao-teorica}
	\input{2-textuais/3-justificativa}

	\input{2-textuais/4-objetivos}
	\input{2-textuais/5-metodologia}
	\input{2-textuais/6-resultados}
	\input{2-textuais/7-conclusao}
	
	\bibliography{3-pos-textuais/referencias}

\end{document}

%% file: lib/preambulo.tex
\usepackage[utf8]{inputenc}                         
\usepackage[T1]{fontenc}                            
\usepackage{graphicx}                               
\usepackage{amsfonts, amssymb, amsmath}             
\usepackage{booktabs}                               
\usepackage{verbatim}                               
\usepackage{multirow, array}                        
\usepackage{indentfirst}                            
\usepackage{listings}                               
\usepackage{longtable}
\usepackage[table,xcdraw]{xcolor}
\usepackage{microtype}                              
\usepackage[portuguese,ruled,lined]{algorithm2e}    
\usepackage{algorithmic}                            
\usepackage{amsgen}
\usepackage{lipsum}                                 
\usepackage{tocloft}                                
\usepackage{etoolbox}                               
\usepackage[nogroupskip,nonumberlist]{glossaries}   

\usepackage[font=singlespacing,format=plain,justification=justified,skip=0pt,singlelinecheck = false]{caption}            

\usepackage[alf, abnt-emphasize=bf, recuo=0cm, abnt-etal-cite=2, abnt-etal-list=0, abnt-etal-text=it]{abntex2cite}  

\usepackage{mathptmx}         
\usepackage{appendix}         
\usepackage{paracol}          
\usepackage{lib/ifcetex}	      
\usepackage{pdfpages}         
\usepackage{amsmath}          

\makeglossaries 
\makeindex      

%% file: 1-pre-textuais/dedicatoria.tex
Aos amigos presentes durante minha formação que tornaram essa etapa tão incrível.

%% file: 1-pre-textuais/agradecimentos.tex
Agradeço primeiramente aos meus orientadores, Dr. César Menezes e Dr. Paulo Parente por me guiarem nesse trabalho e tornarem ele possível.

Ao Instituto Federal de Educação, Ciência e Tecnologia do Ceará - Campus Acaraú por me dar oportunidades de ter experiências tão incríveis. Aos professores e servidores que contribuíram com minha formação, em especial ao professores de física Hygor Piaget, Paulo Santiago, Fellipe Santos, Alex Samyr, Paulo Victor e Luiz Paulo, que juntos ao meu orientador César Menezes foram os professores mais influentes durante minha graduação.

Aos meus amigos, em especial as amizades que cultivei dentro do IFCE por tornarem essa carreira um período fantástico de minha vida. E a todos que me apoiaram durante essa jornada.

%% file: 1-pre-textuais/epigrafe.tex
``Das leis mais simples nascem maravilhas infinitas que se repetem indefinidamente''

\autordaepigrafe{Benoît B. Mandelbrot}


%% file: 1-pre-textuais/resumo.tex
Este trabalho abordou a utilização do movimento browniano geométrico para simulação de preços de ações listadas no índice de \emph{Small Caps} da bolsa de valores brasileira B3 (Brasil, Bolsa, Balcão). Os dados utilizados referem-se ao histórico de preços de janeiro de 2016 a dezembro de 2018. O histórico de preços de 2019 foi utilizado para ser comparado com os preços simulados. Os dados foram importados do banco de dados do \emph{Yahoo Finance} através da linguagem de programação \emph{Python}, e as simulações foram realizadas para cada ação individualmente, e para carteiras formadas com base nos retornos esperados, no risco e no Índice de Sharpe. Os resultados se mostraram melhores para as carteiras de maiores retornos, menores riscos e maiores Índices de Sharpe.

\palavraschave{Econofísica. Movimento Browniano Geométrico. Mecânica Estatística. Small Caps.}

%% file: 1-pre-textuais/abstract.tex
This work addressed the use of the geometric Brownian motion to simulate the prices of shares listed in the Small Caps index of the Brazilian stock exchange B3 (Brazil, Bolsa, Balcão). The data used refer to the price history from January 2016 to December 2018. The price history of 2019 was used to be compared with the simulated prices. The data was imported from the Yahoo Finance database using the Python programming language, and the simulations were performed for each stock individually, and for portfolios formed based on expected returns, risk and the Sharpe Index. The results were better for portfolios with higher returns, lower risks and higher Sharpe Indexes.

\keywords{Econophysics. Geometric Brownian Motion. Statistical Mechanics. Small Caps.}

%% file: 2-textuais/1-introducao.tex
\chapter{Introdução}
\label{cap:introducao}

Quando estudamos ciências naturais, procuramos compreender a natureza em seus aspectos mais gerais e fundamentais. Dentre estas ciências, temos a Física, que estuda a natureza e seus fenômenos buscando explicações para os mesmos. Quando estudamos os diversos fenômenos que ocorrem e formulamos modelos explicativos sobre eles, é possível uma melhor compreensão sobre a natureza.

No século XX iniciou-se o interesse por estudos da dinâmica de sistemas complexos, sistemas que possuem partes que podem interagir de maneira não-linear. Estes sistemas estão presentes na física, biologia, ciências sociais e até no comportamento de mercados financeiros \cite[~p. 99]{gleria2004sistemas}.

Para denominar as aplicações da física estatística à economia, foi cunhado o termo \emph{``econophysics''} por \citeonline{stanley1996anomalous}, ou econofísica, e através dela podemos estudar dados relacionados a finanças, com objetivo de criar modelos explicativos para aqueles tipos de dados, como as ciências naturais sempre fazem.

Na década de 1990, a econofísica começou a circular em meios acadêmicos, e em julho de 1997, foi realizada a primeira conferência sobre o assunto, \emph{``Workshop of Econophysics"} \cite{kondor1999econophysics}. Após isso, essa área foi crescendo, e surgiram livros sobre o assunto, como \emph{An introduction to econophysics} \cite{stanley2000introduction} e \emph{M.Theory of financial risks} \cite{bouchaud2000theory}.

Matemáticos financeiros também utilizam métodos estatísticos para economia, porém, os físicos se concentram em analisar dados experimentais utilizando ferramentas e técnicas que normalmente são aplicadas para sistemas complexos \cite{stanley2001similarities}.

Os estudos sobre física clássica foram responsáveis por grandes impactos na economia. Foi percebido posteriormente que existe analogia nos padrões da natureza com o mercado financeiro. Adam Smith, considerado o pai da economia clássica, toma como exemplo a mecânica celeste e sua enorme quantidade de dados que necessitam serem analisados para que encontremos regularidades para construirmos teorias que expliquem estes dados \cite{smith1880principles}.

\begin{citacao}

Em particular, os conceitos da mecânica foram considerados a ferramenta ideal para a formalização matemática da economia. Alguns economistas como Walras, Jevons, Fisher e Pareto tentaram transferir o formalismo da física para a economia, substituindo pontos materiais por agentes financeiros, fazendo uma analogia entre a energia potencial e o conceito de ``utilidade", e então evoluindo o sistema através de análogos do princípio da mínima ação \cite[~p.3]{favarodinamicas}.

\end{citacao}

A utilização de computadores se tornou muito importante e prática para a análise de dados econômicos, com a capacidade de armazenar um volume enorme de dados, e com dados disponibilizados para todos na Internet, é possível mais pessoas terem acesso a essas informações e fazerem suas próprias análises.

Para estudar as dinâmicas do mercado de ações a partir da econofísica pode-se considerar o mercado como um sistema complexo, e será utilizado o movimento browniano geométrico para a tentativa de simular como os preços de ações poderão se comportar ao longo do tempo baseado em dados históricos. Ao longo do trabalho, a precisão deste método será avaliado.

Este trabalho está organizado da seguinte forma: no Capítulo \ref{chap:referencial-teorico} foi levantado todo o referencial teórico do que foi necessário compreender para a realização do trabalho, no Capítulo \ref{chap:justificativa} foi justificado o motivo da execução e no Capítulo \ref{chap:objetivos} foram apresentados os objetivos propostos para esse estudo.

No Capítulo \ref{chap:metodologia} foi apresentada a metodologia de como o trabalho foi executado, os resultados e discussão sobre eles foram falados no Capítulo \ref{chap:resultados}, e a conclusão foi mostrada no Capítulo \ref{chap:conclusao}.




%% file: 2-textuais/2-fundamentacao-teorica.tex
\chapter{Referencial Teórico}
\label{chap:referencial-teorico}

Neste capítulo será primeiramente abordada uma introdução sobre mecânica estatística, mostrando um dos objetos de estudo dela que são os sistemas complexos, e uma ferramenta aqui escolhida para estudá-los, o movimento browniano geométrico. 

Em seguida serão exploradas algumas propriedades estatísticas para economia que aqui foram utilizadas, tais como retorno, risco, Índice de Sharpe e otimização de carteiras. No fim, foram abordadas as \emph{Small Caps}, que foi o índice de ações aqui explorado.

\section{Mecânica Estatística}\label{sec:mecanica-estatistica}

A mecânica estatística surgiu através da termodinâmica. Enquanto a termodinâmica descreve os efeitos macroscópicos de sistemas formados por grandes números de entes como partículas, células, spins e outros, entes que microscopicamente são descritos através da mecânica clássica ou mecânica quântica, a mecânica estatística é uma teoria probabilística que faz a conexão entre os níveis macroscópico e microscópico \cite[~p. 2]{stariolomecanica}.

Assim como diversas outras áreas da física, o formalismo estatístico se mostrou bem geral, podendo ser utilizado na predição de propriedades bem diversas, como a ocorrência ou não de supercondutividade em um material, a probabilidade de ocorrência de um terremoto, a morfologia típica de uma colônia de células em um tecido vivo ou a evolução dos preços de produtos na bolsa de valores \cite[~p. 3]{stariolomecanica}.

Inicialmente a mecânica estatística foi desenvolvida para predizer propriedades macroscópicas de sistemas em equilíbrio termodinâmico. Porém, por estar relacionada com sistemas dinâmicos, a grande maioria dos sistemas de interesse, físicos ou não, não se encontram em equilíbrio, é o que acontece com o mercado de ações \cite[~p. 3]{stariolomecanica}.

\subsection{Sistemas Complexos}

Os sistemas complexos fazem parte de uma área interdisciplinar que se torna mais importante a cada dia para melhorarmos as explicações sobre os fenômenos que acontecem na natureza. Ela era estudada até o fim do século XIX como algo regido por uma ordem, porém com o passar do tempo foi-se conhecendo alguns fenômenos caóticos \cite[~p. 64]{tsallis2008sistemas}.

As equações de movimento obtidas a partir das leis de Newton possuem uma garantia de que a partir de condições iniciais de um problema, é possível obter a solução para qualquer momento posterior, uma vez que um conjunto de condições iniciais levam a uma única solução e que alterações minúsculas nas condições iniciais produzem pequenas mudanças nas soluções para tempos posteriores. Porém, a validade destas propriedades são estabelecidas somente para regiões limitadas de espaço-tempo \cite[~p. 363]{pires2011evoluccao}.

Quando se observa sistemas de forma mais ampla, é comum nos depararmos com sistemas caóticos como um sistema atmosférico, o formato de litorais e ilhas ou o movimento de conjuntos de astros em sistemas instáveis. A não-linearidade é uma característica de sistemas caóticos, e para se entender o melhor o seu conceito, é necessário definir sistemas não-lineares. ``Em Física um sistema é linear, falando de uma maneira simples, quando o todo é igual à soma de suas partes, e a soma de um conjunto de causas produz uma soma correspondente de efeitos'' \cite[~p.364]{pires2011evoluccao}. Isso é comum quando se vê pequenos problemas isolados, mas não é o único tipo de sistema encontrado na natureza, pois existem os sistemas não-lineares, em que o todo é diferente da soma das partes, e seguem alguns princípios, como o de que equações simples podem levar a movimentos irregulares, equações e sistemas que são apenas ligeiramente diferentes um do outro podem ter para tempo longos, comportamentos bem diferentes, e condições iniciais ligeiramente diferentes podem levar a movimentos radicalmente diferentes \cite[~p.364-365]{pires2011evoluccao}.

Então, definindo um sistema caótico tem-se o seguinte:

\begin{citacao}

Para que um sistema tenha comportamento imprevisível – ou
caótico –, ele deve obedecer a pelo menos três regras: a) ser
dinâmico, ou seja, se alterar à medida que o tempo passa – um
carro se movendo numa estrada; b) ser não linear, isto é, sua
resposta não é proporcional à perturbação – uma simples declaração pode causar uma revolução de estado; c) ser muito
sensível a perturbações mínimas de seu estado, ou seja, uma
alteração desprezível no presente pode causar, no longo prazo,
uma mudança imprevisível – uma leve variação na trajetória
de uma sonda espacial pode levá-la para longe de seu destino \cite[~p.66]{tsallis2008sistemas}.

\end{citacao}

Sistemas ditos complexos foram surgindo em diversas áreas do conhecimento, e até a década de 1980 cada área dava um tratamento específico a seus sistemas complexos \cite[~p.67]{tsallis2008sistemas}. Estes sistemas apareciam por exemplo na física, biologia, sociologia, economia, entre outros.

Ao longo do tempo foi-se percebendo que todos os sistemas complexos possuíam propriedades universais, então estes sistemas se tornaram uma das disciplinas científicas mais interdisciplinares de todas \cite[~p.67]{tsallis2008sistemas}.

Apesar de não existir ainda uma definição única e exata de complexidade, um sistema é considerado mais complexo à medida em que são necessárias mais informações para podermos o descrever. Os sistemas caóticos e complexos têm um aspecto em comum: são não lineares. Só que no caótico a imprevisibilidade é “selvagem”, mais difícil de ser modelada, e no complexo é “civilizada” \cite[~p. 69]{tsallis2008sistemas}. Então é possível estudar melhor um sistema complexo, pois apesar de conter muitas aleatoriedades, eles seguem uma ordem geral.

Com essa interdisciplinaridade, 

\begin{citacao}

Um ramo da teoria dos sistemas complexos que vem
recebendo cada vez mais atenção nos últimos anos é a econofísica que, como o nome sugere, procura compreender o comportamento de mercados financeiros e de outros aspectos da economia \cite[~p. 99]{gleria2004sistemas}.

\end{citacao}

Então, pode-se utilizar estas propriedades universais que os sistemas complexos possuem para um estudo sobre economia ser realizado.

\subsection{Movimento Browniano}

O movimento browniano foi descoberto pelo botânico Robert \citeonline{brown1828xxvii}, que,    ao observar por um microscópio um grão de pólen suspenso na água, notou que seu movimento era irregular, como se tivesse vida própria. Inicialmente Brown pensou ter encontrado uma molécula primitiva para a vida, porém, após mais estudos ele notou que este movimento irregular vinha de princípios físicos, não biológicos.

\begin{citacao}

Nas décadas seguintes, inúmeras tentativas foram
realizadas para desvendar a natureza do movimento
browniano. Experimentos de laboratório mostraram
que o movimento fica mais intenso quando se reduz
a viscosidade do meio ou o tamanho das partículas
brownianas, e também quando se eleva a temperatura
da solução \cite[~p.25]{silva2007quatro}.

\end{citacao}

Segundo \citeonline[~p.25]{silva2007quatro}, foi-se eliminando diversas possíveis causas como atrações ou repulsões entre as partículas suspensas, bolhas temporárias de ar, correntes de convecção no interior da solução, gradientes de temperatura, perturbações mecânicas, até que a partir de 1860 começou-se a tomar forma um novo ponto de vista sobre o movimento browniano, que ele poderia ser devido a colisões com as moléculas do fluido, verificando-se que as trajetórias não possuíam tangentes (então as curvas não eram diferenciáveis) e que o movimento nunca cessava. Porém, só em 1905 que este mistério foi resolvido por Einstein.

O movimento browniano é um movimento aleatório onde a posição em determinado instante de tempo só depende da posição no instante anterior, e não ao resto do caminho, ao serem realizadas várias simulações de caminhos aleatórios é possível observar a distribuição das posições ao longo do tempo, com isso é possível notar que o deslocamento quadrático médio evolui com o tempo \cite[~p. 28]{silva2007quatro}.

Para \citeonline{einstein1905investigations}, o movimento aleatório das partículas suspensas em um fluido se dá devido ao movimento térmico dos átomos que constituem o líquido, e para o movimento browniano, o conceito de velocidade não fazia mais sentido já que o espaço percorrido não é proporcional ao tempo, e sim a raiz quadrada dele.

Mais tarde, em 1923, o movimento browniano foi formalizado matematicamente por Norbert Wiener. Por isso, este movimento também é conhecido como Processo de Wiener \cite[~p.147]{luizteoremas}.

\begin{figure}[!h] 
   	    \captionsetup{width=13cm}
		\Caption{\label{fig:wiener} Processo de Wiener com distribuição de média 0, desvio padrão 1, e 1000 passos em 100 unidades de tempo}
		\ifcefig{}{
			\includegraphics[width=13cm]{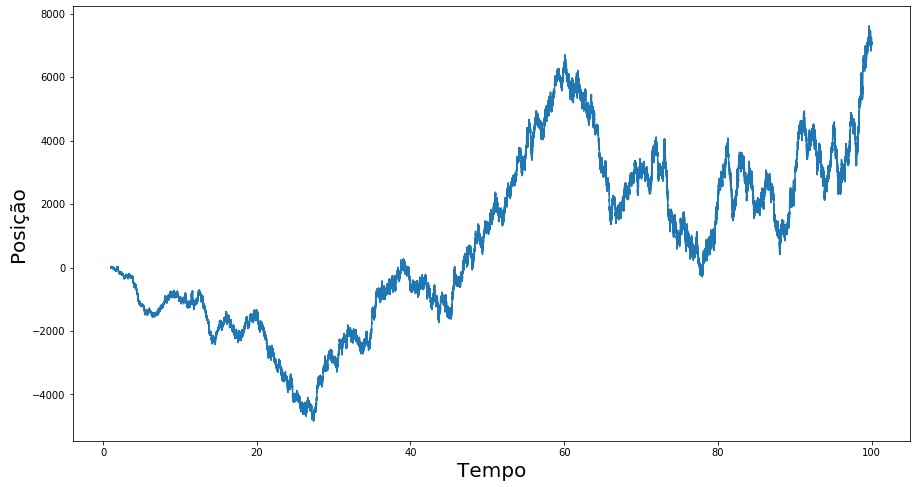}
		}{
			\Fonte{Elaborada pelo autor (2020).}
		}	
	\end{figure}

Um processo de Wiener é descrito pela equação:
\begin{equation}
    dW(t) = \epsilon_t\sqrt{dt},
    \label{eq:wiener}
\end{equation}

\noindent onde $dW$ é cada passo do movimento, $t$ é o tempo e $\epsilon_t$ é uma variável aleatória com distribuição de média 0 e desvio padrão 1 \cite[~p.25]{priscilla2011}. A Figura \ref{fig:wiener} mostra um exemplo de movimento com passos governados por um processo de Wiener.

\subsubsection{Movimento Browniano Geométrico}

Segundo \citeonline[~p. 195]{levadaconsideraccoes}, o processo de interação entre compradores e vendedores que fazem variar o preço de uma opção\footnote{Opção é um tipo de contrato que dá ao investidor o direito de comprar ou vender determinado ativo em data e preço preestabelecidos.} é regido por um movimento browniano. Então, em aproximação, pode-se utilizar a matemática do movimento browniano para simular ``caminhos'' para o preço e tentar prever o valor de uma ação em determinado instante de tempo.

Para modelar o comportamento do preço de ações ao longo do tempo pode-se utilizar o movimento browniano geométrico \cite[~p. 5]{brigo2007stochastic}. Ele satisfaz a seguinte equação diferencial estocástica:
\begin{equation}
    dS(t) = \mu S(t)dt + \sigma S(t)dW(t),
    \label{eq:mbg}
\end{equation}

\noindent onde $S(t)$ é a variação percentual do preço da ação ao longo do tempo, $\mu$ e $\sigma$ são respectivamente a média e o desvio padrão dos retornos, e $W(t)$ é um processo de Wiener, definido pela Equação \ref{eq:wiener}, então o primeiro termo modela as tendências determinísticas, enquanto o segundo modela o comportamento aleatório do movimento browniano geométrico. A Equação \ref{eq:mbg} tem a seguinte solução:
\begin{equation}
    S(t) = S(0)e^{(\mu - \frac{1}{2}\sigma^2)t + \sigma W(t)}.
    \label{eq:sol_mbg}
\end{equation}

\begin{figure}[!h] 
   	    \captionsetup{width=11.5cm}
		\Caption{\label{fig:mbg} Movimento Browniano Geométrico com $\mu = 0,0004$ e $\sigma = 0,01$}
		\ifcefig{}{
			\includegraphics[width=11.5cm]{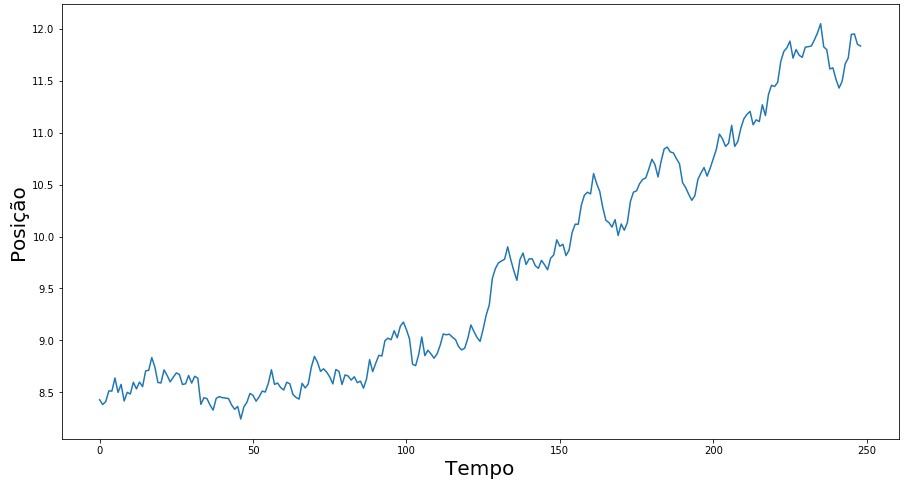}
		}{
			\Fonte{Elaborada pelo autor (2020).}
		}	
	\end{figure}

\noindent A partir dos resultados da Equação \ref{eq:sol_mbg}, é montada a trajetória do movimento browniano geométrico. Figura \ref{fig:mbg} mostra um exemplo de trajetória.

A Equação \ref{eq:sol_mbg} será utilizada na simulação de preços das ações, com $\mu$ e $\sigma$ baseados no histórico dos dados.

\section{Propriedades Estatísticas Para Economia}\label{sec:propriedades-estatisticas-economia}

Quando se estuda sobre ações da bolsas de valores, pode-se utilizar diversos métodos estatísticos para analisar ações ou carteiras de ações com base em seus históricos. Os resultados obtidos com a utilização de tais métodos são de grande importância para os investidores.
\subsection{Preço das Ações}

\begin{figure}[h!] 
   	    \captionsetup{width=15cm}
		\Caption{\label{fig:stocks} Histórico de preços diários de Banco Pan (BPAN4), Even (EVEN3), Fleury (FLRY3) e Sinqia (SQIA3) de 04/01/2016 até 30/12/2019}
		\ifcefig{}{
			\includegraphics[width=15cm]{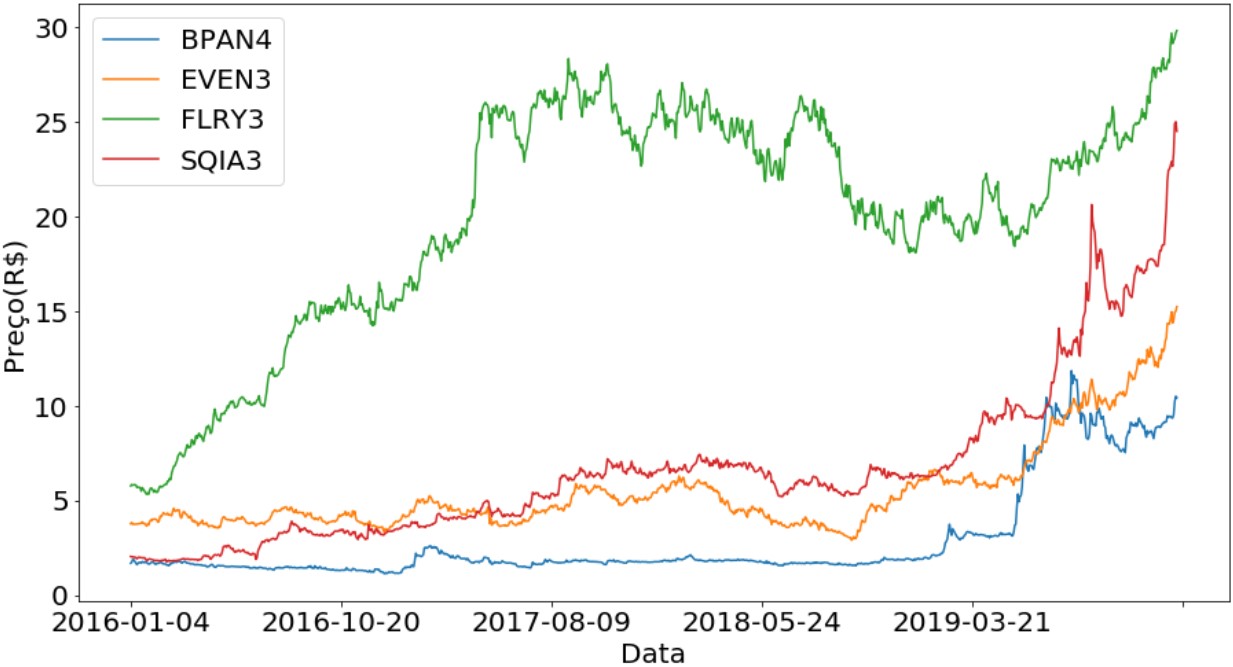}
		}{
			\Fonte{Elaborada pelo autor (2020).}
		}	
	\end{figure}

O primeiro passo para se analisar o histórico das ações consiste em obter essas informações. A linguagem de programação \emph{Python} é muito utilizada para extração e análise estatística de dados. Cada ação tem um código para representá-la. Com esse código, através do banco de dados do \emph{Yahoo Finances}\footnote{https://finance.yahoo.com/}, por exemplo, é possível importar os preços das ações em cada dia útil dentro de um período determinado. A Figura \ref{fig:stocks} ilustra alguns exemplos extraídos desse banco de dados, onde pode-se observar o histórico de preços das ações do Banco Pan (BPAN4), Even (EVEN3), Fleury (FLRY3) e Sinqia (SQIA3).

\begin{figure}[h!] 
   	    \captionsetup{width=15cm}
		\Caption{\label{fig:normalized} Histórico de preços diários normalizados por 100 de Banco Pan (BPAN4), Even (EVEN3), Fleury (FLRY3) e Sinqia (SQIA3) de 04/01/2016 até 30/12/2019}
		\ifcefig{}{
			\includegraphics[width=15cm]{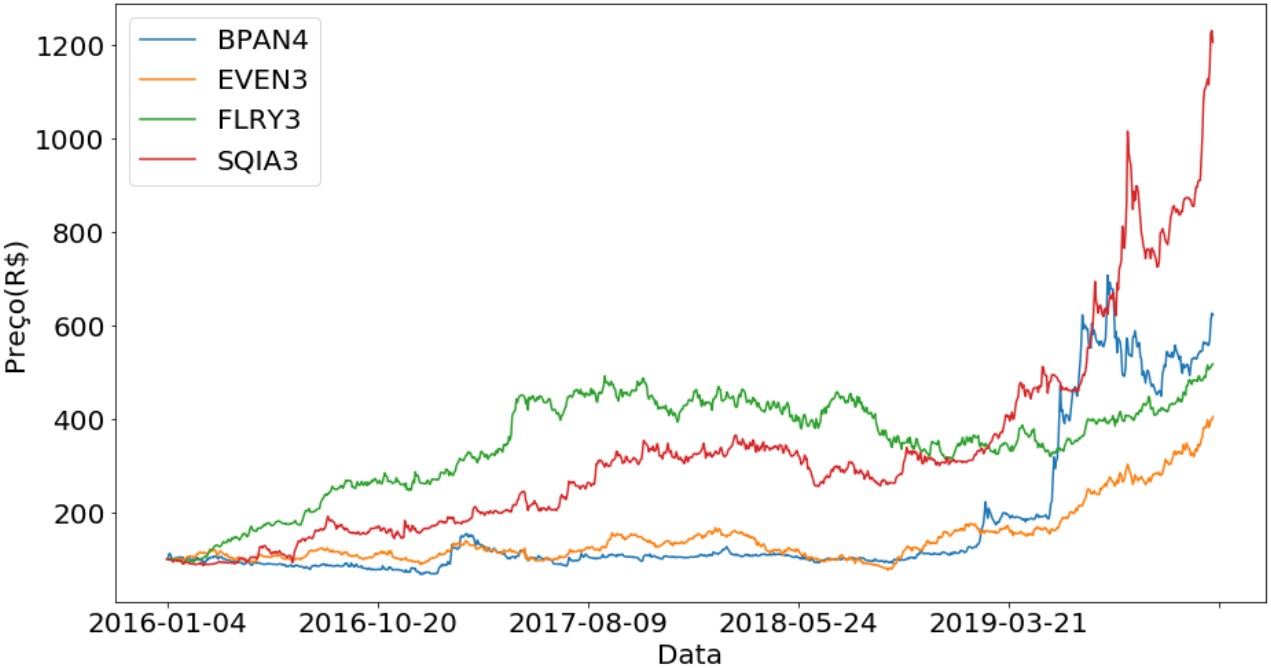}
		}{
			\Fonte{Elaborada pelo autor (2020).}
		}	
	\end{figure}

A diferença entre os preços de cada empresa pode dificultar a visualização dos dados. Para eliminar essa dificuldade, pode-se normalizar os preços por 100, assim, considera-se que o preço de cada ação se inicia em 100 para se visualizar melhor as evoluções. Para isso, basta dividir cada valor de ação por seu valor inicial, e multiplicar por 100. O resultado é mostrado na Figura \ref{fig:normalized}.

Com o gráfico da Figura \ref{fig:normalized}, pode-se visualizar mais facilmente como foi a evolução dos preços de cada ação e assim pode-se compará-las entre si. Em uma carteira com R\$ 100,00 investidos em cada uma dessas ações na data inicial do gráfico, o valor investido em cada uma evoluiria da forma mostrada nele.  Nesses gráficos é possível ter uma primeira ideia de qual ação mais cresceu seu preço, porém, apesar de ser a primeira visão para comparação, estes gráficos não são o suficiente para uma boa análise, pois existem outras informações que podem ser retiradas desse histórico com grande relevância.

\subsection{Retorno}

Quando se precisa calcular quanto uma ação retornou em um dia, é necessário tomar quanto o ativo variou do dia anterior ($P_0$) para o dia atual ($P_1$), e dividir pelo preço do dia anterior ($P_0$) \cite{vieira_analisando_2019}, conforme mostra a Equação \ref{eq:retorno_linear}.
\begin{equation}
    \label{eq:retorno_linear}
    R_s = \frac{P_1 - P_0}{P_0},
\end{equation}

\noindent ou
\begin{equation}
    R_s = \frac{P_1}{P_0} - 1.
\end{equation}

\noindent O retorno calculado desta forma é conhecido como Retorno Simples ($R_s$). Na prática, quando se calcula um retorno diário, os valores do dia atual e do anterior costumam ser bem próximos, desta forma $|R_s| << 1$, assim, é possível fazer uma boa aproximação do retorno para $ln\left(\frac{P_1}{P_0}\right)$ através de uma série de Taylor, pois podemos expandir $ln\left(\frac{P_1}{P_0}\right)$ nas vizinhanças de $\frac{P_1}{P_0} = 1$ da seguinte forma:
\begin{equation}
    ln{\left(\frac{P_1}{P_0}\right)} = ln(1) + \frac{\left(\frac{P_1}{P_0} - 1\right)}{1!} - \frac{\left(\frac{P_1}{P_0} - 1\right)^2}{2!} + ...
\end{equation}

Se a série for truncada no termo de primeira ordem, tem-se que $ln\left(\frac{P_1}{P_0}\right) \approx 0 + \left(\frac{P_1}{P_0} - 1\right)$, chamamos essa aproximação de Retorno Logarítmico ($R_{log}$), então:
\begin{equation}
    \label{eq:retorno_log}
    R_{log} = \ln{\left(\frac{P_1}{P_0}\right)}
\end{equation}

Convém utilizar retorno logarítmico devido as propriedades do logaritmo que permitem que retornos acumulados sejam somados, isso permite que ao se calcular o retorno em um período maior, é possível simplesmente somar os retornos logarítmicos diários desse período, o que não pode ser feito com o retorno simples.

\begin{figure}[!h] 
   	    \captionsetup{width=13cm}
		\Caption{\label{fig:simples} Retornos simples diários de Banco Pan (BPAN4), Even (EVEN3), Fleury (FLRY3) e Sinqia (SQIA3) de 04/01/2016 até 30/12/2019}
		\ifcefig{}{
			\includegraphics[width=13cm]{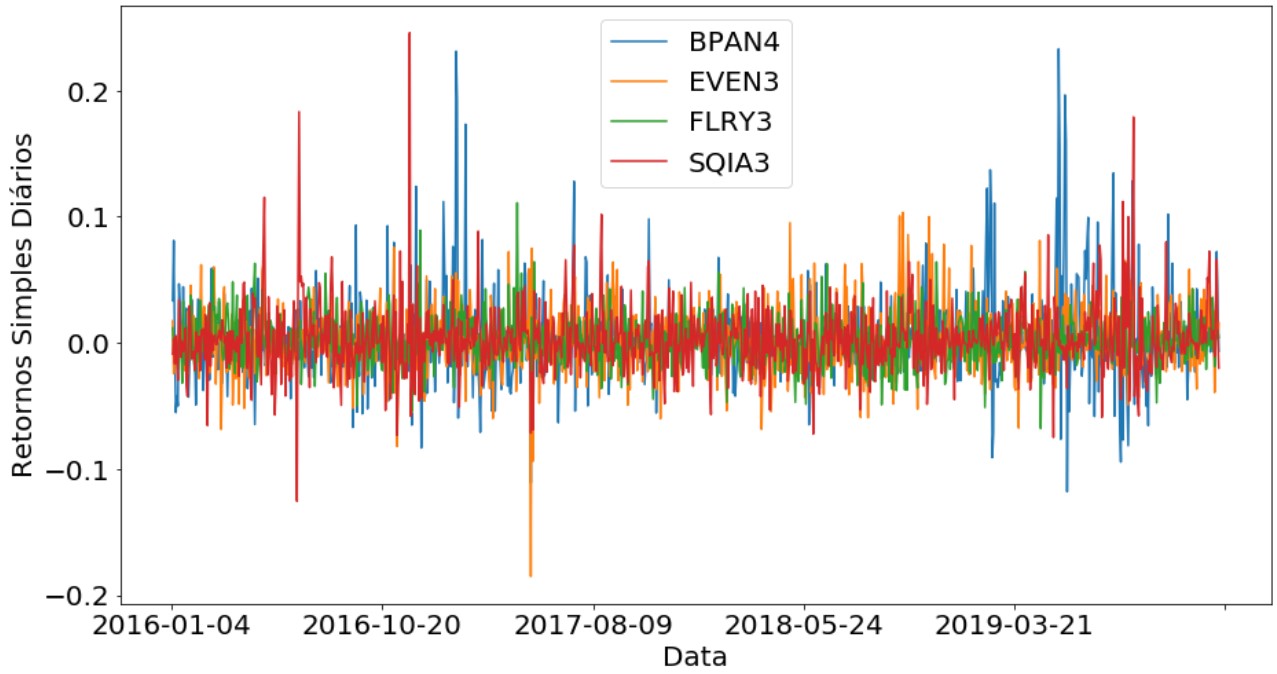}
		}{
			\Fonte{Elaborada pelo autor (2020).}
		}	
	\end{figure}

\begin{figure}[!h] 
   	    \captionsetup{width=13cm}
		\Caption{\label{fig:log} Retornos logarítmicos diários de Banco Pan (BPAN4), Even (EVEN3), Fleury (FLRY3) e Sinqia (SQIA3) de 04/01/2016 até 30/12/2019}
		\ifcefig{}{
			\includegraphics[width=13cm]{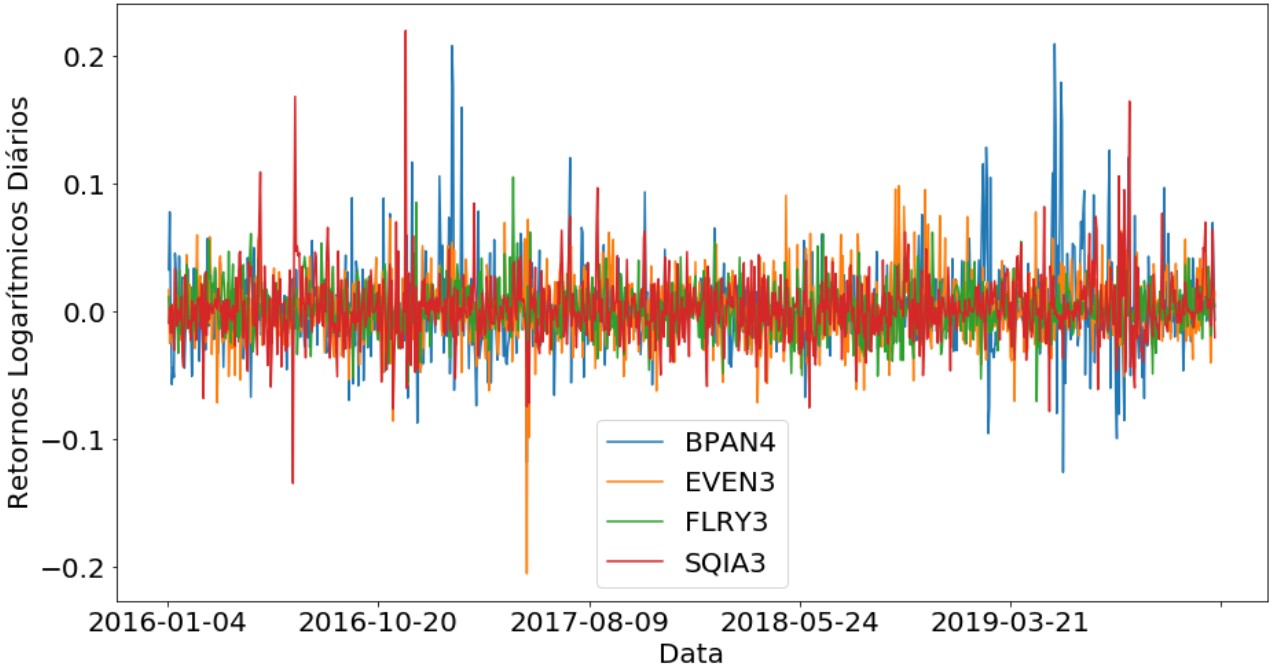}
		}{
			\Fonte{Elaborada pelo autor (2020).}
		}	
	\end{figure}

O gráfico dos retornos simples das ações analisadas anteriormente podem ser vistos na Figura \ref{fig:simples}, enquanto o gráfico dos retornos logarítmicos são mostrados na Figura \ref{fig:log}, são gráficos com grande semelhança, já que o retorno logarítmico é uma boa aproximação para o retorno simples, então visualmente é difícil distingui-los, já que seus valores não diferem muito. Estes retornos são percentuais, então cada informação acima de 0 representa um aumento no preço da ação, e cada uma abaixo de 0, significa que o preço da ação diminuiu. Esta informação é importante na análise.

Se for tomada a média dos retornos diários, e multiplicada por 252, que é a quantidade aproximada de dias úteis em um ano, tem-se o retorno logarítmico médio anual de cada ação. Fazendo isso é possível concluir que o retorno médio anual para BPAN4 é  aproximadamente 46,3\%, para EVEN3, 35,4\%, para FLRY3, 41,7\% e para SQIA3 63,1\%, considerando o período ente 04/01/2016 a 30/12/2019.

\subsection{Risco}

Outro assunto importante para a análise de ações é o risco, pois os investidores querem ter um retorno alto, porém tem que se levar em consideração o risco de perder dinheiro, o que leva alguns deles a tomarem atitudes mais conservadoras. Volatilidade e risco são sinônimos nesse contexto, e risco é definido como o desvio padrão ($\sigma$) dos retornos diários dos ativos.

``Conceitos de estatística como variância e desvio padrão podem nos ajudar quando desejamos quantificar o risco de um ativo. É esperado que uma ação volátil possa desviar muito do seu valor médio, para mais ou para menos'' \cite{vieira_analisando_2019}.

Então, ao quantificar a volatilidade anual de um ativo se utiliza o cálculo do desvio padrão dos retornos diários multiplicado por $\sqrt{252}$, que é representado pela Equação \ref{desvio_padrao}, onde $x_i$ representa o valor do retorno em cada dia, $\bar x$ a média dos valores para $x$, e $n$ é o número de dados, assim obtemos o risco anualizado ($\sigma_ i$) do ativo.
\begin{equation}
    \sigma_i = \sqrt{\sum_i{\frac{(x_i - \bar x)^2}{n-1}}} \cdot \sqrt{252}
    \label{desvio_padrao}
\end{equation}

Para as ações que estão sendo analisadas no exemplo ilustrativo, foi encontrado o risco de aproximadamente 51,4\% para BPAN4, 40,9\% para EVEN3, 30,3\% para FLRY3 e 41,5\% para SQIA3.

Também existem outras formas de quantificar o risco de uma ação, como o coeficiente Beta, que relaciona a ação com o mercado em que ela está inserido, e para carteiras algumas metodologias utilizam a covariância entre as ações que a compõem.

\subsection{Índice de Sharpe}

\newacronym{IS}{IS}{Índice de Sharpe}

Como foi visto, dois fatores importantes que um investidor deve analisar antes de comprar uma ação são o retorno e o risco. Ao escolher uma ação, é preferível que ela tenha o maior retorno possível com menor risco.

Para avaliar o retorno e risco de um investimento, Willian Sharpe (Nobel de economia em 1990) criou um indicador chamado \gls{IS} \cite{carvalho_o_2011}. Este índice calcula o retorno excedente de uma ação ou carteira em relação a uma taxa livre de risco, e é baseado na taxa de retorno anualizado ($R_i$), que pode ser o logarítmico, risco anualizado ($\sigma i$) e a taxa de retorno livre de risco do país ($R_f$), que pode ser definida como algum investimento conservador no país. Para o Brasil pode-se utilizar a taxa CDI ou a taxa SELIC, essas taxas são utilizadas como \emph{benchmark}, são referências de mercado para que um investidor possa acompanhar o desempenho de seu investimento. Assim, o \gls{IS} pode ser calculado com a Equação \ref{eq:sharpe}.
\begin{equation}
    IS = \frac{R_i - R_f}{\sigma_i}
    \label{eq:sharpe}
\end{equation}

Considerando a taxa de retorno livre de risco como o CDI\footnote{Taxa consultada no dia 19/09/2020 no site do Banco Central do Brasil igual a 1,9\%}, para as ações exemplificadas BPAN4, EVEN3, FLRY3 e SQIA3 foram encontrados os Índices de Sharpe de 0,864; 0,819; 1,314 e 1,475 respectivamente.

\subsection{Otimização de Carteira de Ações}\label{sec:otimizacao}

Ao se montar uma carteira de ações, é necessário analisar uma boa distribuição de investimento entre essas ações, pois a diversificação é uma boa forma de melhorar o rendimento de uma carteira \cite{zanini2005teorias}.

Ao diversificar a carteira, de acordo com a proporção de investimento em cada ação, se modifica os retornos e os riscos esperados dela, então com uma boa análise é possível maximizar o retorno  para um dado risco que se deseja tomar. \citeonline{doi:10.1111/j.1540-6261.1952.tb01525.x} e \citeonline{sharpe1963simplified} trabalharam em modelos para otimizar carteiras de ações com balanceamento do quanto foi investido em cada uma.

Uma forma possível para essa otimização, baseada nos modelos propostos por \citeonline{doi:10.1111/j.1540-6261.1952.tb01525.x} e \citeonline{sharpe1963simplified}, é uma simulação massiva de carteiras com distribuições aleatórias de investimento em cada ação escolhida, pode-se fazer o cálculo do \gls{IS} para cada carteira e achar a de maior índice.

\begin{figure}[h!] 
   	    \captionsetup{width=12.5cm}
		\Caption{\label{fig:carteiras} Carteiras com distribuições aleatórias de investimentos para BPAN4, EVEN3, FLRY3 e SQIA3 baseadas em dados de 04/01/2016 até 30/12/2019}
		\ifcefig{}{
			\includegraphics[width=12.5cm]{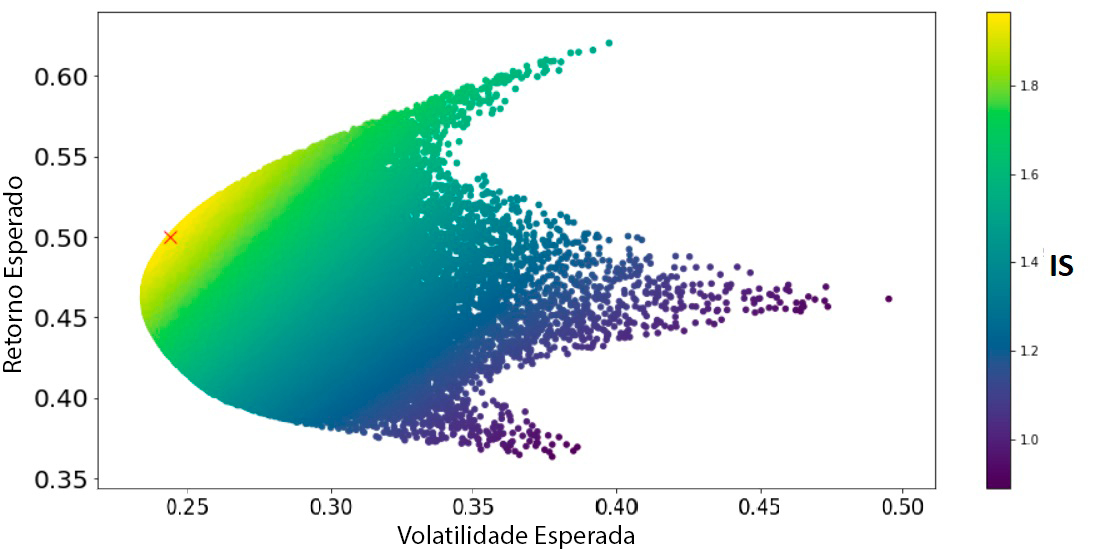}
		}{
			\Fonte{Elaborada pelo autor (2020).}
		}	
	\end{figure}

Uma simulação foi feita com as ações aqui exemplificadas. Ao fazer essa simulação, pode-se construir um gráfico do retorno pelo risco de cada carteira. No gráfico da Figura \ref{fig:carteiras} foram simuladas cem mil carteiras.

As cores dos pontos do gráfico estão variando com o \gls{IS} de cada carteira (índices maiores estão mais próximos do amarelo), e o ``X'' vermelho marca a carteira com maior \gls{IS} entre todas as simuladas. Com isso, foi possível deduzir que de acordo com esse parâmetro, a carteira mais eficiente para ser montada com essas ações é uma carteira com aproximadamente 15\% do investimento em BPAN4; 4\% em EVEN3; 44\% em FLRY3 e 37\% em SQIA3.

\section{\emph{Small Caps}}

Atualmente, na bolsa de valores brasileira, existem mais de 350 ações sendo negociadas \cite{elias_354_2020}. Com o desenvolvimento da bolsa, houve um período em que foram criados índices de segmentação das empresas em alguns grupos, dentre esses índices, está o de \emph{Small Caps}, desenvolvido em 2008, em meio a uma crise financeira internacional com desvalorização superior a 50\% naquele ano, porém, no ano seguinte teve valorização de 137,5\% \cite[~p.5]{drumond2012analise}.

No geral, enquadram-se como \emph{Small Caps} empresas com valor de mercado entre 300 milhões e 2 bilhões de reais, valores que podem sofrer alterações conforme a cotação do dólar \cite{aguillar_small_2019}.

Quando comparadas a outras empresas listadas na bolsa de valores, \emph{Small Caps} são consideradas empresas pequenas, e diferente das grandes empresas, elas costumam apresentar maior risco e maior retorno, sendo consideradas menos conservadoras. Essas empresas estão ingressando no mercado e podem ter um futuro promissor pela frente, facilitando para o investidor a possibilidade de maior retorno em curto prazo \cite{aguillar_small_2019}.

\citeonline{drumond2012analise}, em um estudo, analisou carteiras otimizadas composta por ações do índice de \emph{Small Caps} e por ações de maior capitalização no mercado e concluiu que as carteiras de \emph{Small Caps} tiveram maior desempenho em janelas de um, três e cinco anos.

%% file: 2-textuais/3-justificativa.tex
\chapter{Justificativa}
\label{chap:justificativa}

É importante verificar os ambientes de atuação da física, pois muitos não conhecem essa ponte que existe ligando a mecânica estatística ao comportamento de mercados de ações através do movimento browniano geométrico.

Existem trabalhos que abordam aplicações do movimento browniano geométrico como realizados por \citeonline{postali2006geometric}, \citeonline{gatheral2011optimal} e \citeonline{gerber2003geometric}, porém, foi notado que poucos testam sua precisão. Além disso, frequentemente o movimento browniano geométrico é utilizado para simular ações individualmente.

Um trabalho onde foram feitos testes de precisão na simulação do movimento browniano geométrico para carteiras de ações foi o de \citeonline{reddy2016simulating}, que escolheram ações listadas dentre as mais negociadas da Austrália, enquanto neste trabalho foram escolhidas \emph{Small Caps} brasileiras e, além de as carteiras simuladas serem organizadas por retorno e volatilidade, também foram simuladas aqui carteiras otimizadas com base nos modelos de \citeonline{doi:10.1111/j.1540-6261.1952.tb01525.x} e \citeonline{sharpe1963simplified}.

Há uma lacuna na literatura de trabalhos analisando simulações de preços de \emph{Small Caps}, normalmente se analisam índices macros no mercado como o Ibovespa, como foi feito por \citeonline{bessada2005generalizaccoes}. As \emph{Small Caps} são interessantes por terem maior potencial de crescimento e maior risco, com isso, alguns investidores conseguem em curto período grandes retornos investindo nelas, diferente das empresas de índices maiores, que são mais conservadoras, onde só é esperado bom retorno a longo prazo.

%% file: 2-textuais/4-objetivos.tex
\chapter{Objetivos}
\label{chap:objetivos}

\section{Objetivo Geral}

\begin{itemize}
    \item Simular os preços das ações e das carteiras de ações do índice \emph{Small Caps} utilizando o movimento browniano geométrico e compará-las com os preços reais.
\end{itemize}

\section{Objetivos Específicos}

\begin{itemize}
    \item Simular o preço das ações individuais do índice \emph{Small Caps} e comparar com os preços reais;
    \item Simular o preço das carteiras de ações do índice \emph{Small Caps} formadas por retorno, risco e \gls{IS} e comparar com os preços reais.
\end{itemize}

%% file: 2-textuais/5-metodologia.tex
\chapter{Metodologia}
\label{chap:metodologia}

\section{Dados Utilizados}

Foram utilizados os dados do histórico de preços de ações listadas na bolsa de valores brasileira do índice \emph{Small Caps}. Para a preparação da simulação foram utilizados os dados do início de 2016 ao fim de 2018 e para comparação entre os preços reais e simulados foram utilizados os dados do início ao fim de 2019. Algumas empresas não tinham capital aberto em 2016, então estas foram excluídas dos dados, restando 78 ações que estão representadas na Tabela \ref{tab:acoes}.

\begin{table}[!h]
\captionsetup{width=15.7cm}
\centering
\small
\caption{Ações utilizadas para a simulação}
\label{tab:acoes}
\begin{tabular}{|c|c|c|c|c|c|}
\hline
\rowcolor[HTML]{C0C0C0} 
Acão         & Código & Ação         & Código & Ação         & Código \\ \hline
ABC   BRASIL & ABCB4  & EZTEC        & EZTC3  & MARCOPOLO    & POMO4  \\ \hline
ALIANSCSONAE & ALSO3  & FLEURY       & FLRY3  & POSITIVO TEC & POSI3  \\ \hline
ALUPAR       & ALUP11 & GAFISA       & GFSA3  & PETRORIO     & PRIO3  \\ \hline
LOJAS MARISA & AMAR3  & GERDAU MET   & GOAU4  & QUALICORP    & QUAL3  \\ \hline
ANIMA        & ANIM3  & GOL          & GOLL4  & RANDON PART  & RAPT4  \\ \hline
AREZZO CO    & ARZZ3  & GRENDENE     & GRND3  & COSAN LOG    & RLOG3  \\ \hline
MINERVA      & BEEF3  & GUARARAPES   & GUAR3  & SANEPAR      & SAPR4  \\ \hline
BANCO PAN    & BPAN4  & HELBOR       & HBOR3  & SANEPAR      & SEER3  \\ \hline
BRADESPAR    & BRAP4  & CIA HERING   & HGTX3  & SER EDUCA    & SLCE3  \\ \hline
BR MALLS PAR & BRML3  & IGUATEMI     & IGTA3  & SLC AGRICOLA & SMLS3  \\ \hline
BR PROPERT   & BRPR3  & JHSF PART    & JHSF3  & SMILES       & SMTO3  \\ \hline
BANRISUL     & BRSR6  & JSL          & JSLG3  & SINQIA       & SQIA3  \\ \hline
CENTAURO     & CESP6  & LOCAMERICA   & LCAM3  & SANTOS BRP   & STBP3  \\ \hline
COGNA ON     & COGN3  & METAL LEVE   & LEVE3  & TAESA        & TAEE11 \\ \hline
COPASA       & CSMG3  & LIGHT S/A    & LIGT3  & TAURUS ARMAS & TASA4  \\ \hline
CVC BRASIL   & CVCB3  & LINX         & LINX3  & TECNISA      & TCSA3  \\ \hline
CYRELA REALT & CYRE3  & LOG COM PROP & LOGN3  & TENDA        & TEND3  \\ \hline
DIRECIONAL   & DIRR3  & LOPES BRASIL & LPSB3  & TEGMA        & TGMA3  \\ \hline
DOMMO        & DMMO3  & M.DIASBRANCO & MDIA3  & AES TIETE E  & TIET11 \\ \hline
DURATEX      & DTEX3  & IMC S/A      & MEAL3  & TRISUL       & TRIS3  \\ \hline
ECORODOVIAS  & ECOR3  & MILLS        & MILS3  & TUPY         & TUPY3  \\ \hline
EMBRAER      & EMBR3  & MARFRIG      & MRFG3  & USIMINAS     & USIM5  \\ \hline
ENAUTA PART  & ENAT3  & MRV          & MRVE3  & VALID        & VLID3  \\ \hline
ENERGIAS BR  & ENBR3  & MULTIPLAN    & MULT3  & VULCABRAS    & VULC3  \\ \hline
ENEVA        & ENEV3  & IOCHP-MAXION & MYPK3  & WIZ S.A.     & WIZS3  \\ \hline
EVEN         & EVEN3  & ODONTOPREV   & ODPV3  & YDUQS PART   & YDUQ3  \\ \hline
\end{tabular}
\Fonte{B3, 2020.}
\end{table}

\section{Organização dos dados}\label{sec:organizacao}

Para a organização dos dados, em cada ação foram calculadas as médias do retornos logaritmos e dos riscos diários e os \gls{IS}.

Em seguida, as carteiras que seriam simuladas foram formadas. A primeira categoria de carteiras foi organizada por retorno esperado, as ações foram ordenadas do maior para o menor retorno, e separadas em 6 carteiras de 13 ações cada, e investimento igualmente dividido entre elas, sendo considerado R\$ 100,00 investidos em cada ação, sendo a Carteira 1 com as 13 ações de maior retorno, e a Carteira 6 com as 13 ações de menor retorno. As carteiras ficaram organizadas de acordo com a Tabela \ref{tab:cart_retorno}.

\begin{table}[!h]
\centering
\small
\captionsetup{width=14.75cm}
\caption{Carteiras organizadas por retorno}
\label{tab:cart_retorno}
\begin{tabular}{|c|c|c|c|c|c|}
\hline
\rowcolor[HTML]{C0C0C0} 
\begin{tabular}[c]{@{}c@{}}Carteira 1\\ (Maiores retornos)\end{tabular} & Carteira 2 & Carteira 3 & Carteira 4 & Carteira 5 & \begin{tabular}[c]{@{}c@{}}Carteira 6\\ (Menores retornos)\end{tabular} \\ \hline
LCAM3      & RAPT4      & ABCB4      & MULT3      & ALSO3      & BPAN4      \\ \hline
SAPR4      & TRIS3      & POMO4      & ENAT3      & ENBR3      & AMAR3      \\ \hline
GOLL4      & HGTX3      & DIRR3      & GRND3      & EVEN3      & COGN3      \\ \hline
PRIO3      & VULC3      & CYRE3      & MRVE3      & POSI3      & BRPR3      \\ \hline
BRAP4      & SLCE3      & IGTA3      & LIGT3      & ANIM3      & TIET11     \\ \hline
TGMA3      & TASA4      & LINX3      & BRML3      & LOGN3      & GFSA3      \\ \hline
USIM5      & TEND3      & DTEX3      & CESP6      & SMTO3      & MRFG3      \\ \hline
CVCB3      & RLOG3      & SEER3      & ALUP11     & TUPY3      & WIZS3      \\ \hline
CSMG3      & FLRY3      & ECOR3      & MEAL3      & ENEV3      & EMBR3      \\ \hline
GUAR3      & SQIA3      & MYPK3      & MILS3      & JSLG3      & TCSA3      \\ \hline
BRSR6      & ARZZ3      & YDUQ3      & ODPV3      & LEVE3      & VLID3      \\ \hline
GOAU4      & EZTC3      & MDIA3      & JHSF3      & HBOR3      & BEEF3      \\ \hline
STBP3      & LPSB3      & TAEE11     & SMLS3      & QUAL3      & DMMO3      \\ \hline
\end{tabular}
\Fonte{Elaborada pelo autor (2020).}
\end{table}

A segunda categoria de carteiras foi ordenada por risco, sendo também formadas 6 carteiras de 13 ações cada, com R\$ 100,00 investidos em cada ação, mas dessa vez com a Carteira 1 com as 13 ações com maior risco, e a Carteira 6 com as 13 ações de menor risco. A Tabela \ref{tab:cart_risco} ilustra a organização das carteiras.

\begin{table}[!h]
\centering
\small
\captionsetup{width=14cm}
\caption{Carteiras organizadas por risco}
\label{tab:cart_risco}
\begin{tabular}{|c|c|c|c|c|c|}
\hline
\rowcolor[HTML]{C0C0C0} 
\begin{tabular}[c]{@{}c@{}}Carteira 1\\ (Maiores riscos)\end{tabular} & Carteira 2 & Carteira 3 & Carteira 4 & Carteira 5 & \begin{tabular}[c]{@{}c@{}}Carteira 6\\ (Menores riscos)\end{tabular} \\ \hline
DMMO3      & POSI3      & RLOG3      & POMO4      & EZTC3      & LEVE3      \\ \hline
GOLL4      & JHSF3      & QUAL3      & LCAM3      & CESP6      & FLRY3      \\ \hline
TASA4      & ENAT3      & BRSR6      & JSLG3      & GUAR3      & MRVE3      \\ \hline
SAPR4      & SMLS3      & BPAN4      & DTEX3      & SLCE3      & ODPV3      \\ \hline
VULC3      & LPSB3      & ANIM3      & ECOR3      & CVCB3      & ENBR3      \\ \hline
STBP3      & ENEV3      & TGMA3      & DIRR3      & LINX3      & IGTA3      \\ \hline
PRIO3      & BRAP4      & COGN3      & ARZZ3      & BRPR3      & GRND3      \\ \hline
LOGN3      & RAPT4      & MYPK3      & CYRE3      & BRML3      & ALSO3      \\ \hline
USIM5      & AMAR3      & TCSA3      & CSMG3      & BEEF3      & MULT3      \\ \hline
HGTX3      & LIGT3      & EVEN3      & MRFG3      & MDIA3      & TAEE11     \\ \hline
GOAU4      & GFSA3      & WIZS3      & EMBR3      & TUPY3      & SMTO3      \\ \hline
MILS3      & YDUQ3      & TRIS3      & MEAL3      & TEND3      & TIET11     \\ \hline
HBOR3      & SEER3      & SQIA3      & VLID3      & ABCB4      & ALUP11     \\ \hline
\end{tabular}
\Fonte{Elaborada pelo autor (2020).}
\end{table}

Na terceira categoria de carteiras, primeiramente elas foram ordenadas de acordo com seus \gls{IS}, novamente em 6 carteiras de 13 ações cadas, da forma mostrada na Tabela \ref{tab:cart_sharpe}. Mas, desta vez, a distribuição de investimento não foi igual para cada ação, cada carteira foi otimizada com base nos métodos formulados por \citeonline{doi:10.1111/j.1540-6261.1952.tb01525.x} e \citeonline{sharpe1963simplified} para que cada carteira tenha o maior \gls{IS} possível com as ações que ela contém.

\begin{table}[!h]
\centering
\small
\captionsetup{width=14.2cm}
\caption{Carteiras organizadas por IS}
\label{tab:cart_sharpe}
\begin{tabular}{|c|c|c|c|c|c|}
\hline
\rowcolor[HTML]{C0C0C0} 
\begin{tabular}[c]{@{}c@{}}Carteira 1\\ (Maiores Índices \\ de Sharpe)\end{tabular} & Carteira 2 & Carteira 3 & Carteira 4 & Carteira 5 & \begin{tabular}[c]{@{}c@{}}Carteira 6\\ (Menores Índices \\ de Sharpe)\end{tabular} \\ \hline
LCAM3      & IGTA3      & GRND3      & VULC3      & SMTO3      & HBOR3      \\ \hline
CVCB3      & GOLL4      & DIRR3      & BRML3      & MILS3      & AMAR3      \\ \hline
TGMA3      & ARZZ3      & POMO4      & SEER3      & SMLS3      & COGN3      \\ \hline
GUAR3      & ABCB4      & HGTX3      & TASA4      & JHSF3      & BRPR3      \\ \hline
CSMG3      & RAPT4      & CYRE3      & ALSO3      & ANIM3      & TIET11     \\ \hline
BRAP4      & EZTC3      & MDIA3      & ODPV3      & POSI3      & GFSA3      \\ \hline
SAPR4      & USIM5      & ALUP11     & ENBR3      & TUPY3      & MRFG3      \\ \hline
FLRY3      & SQIA3      & LPSB3      & YDUQ3      & LEVE3      & WIZS3      \\ \hline
TEND3      & TAEE11     & DTEX3      & CESP6      & JSLG3      & EMBR3      \\ \hline
SLCE3      & RLOG3      & MRVE3      & MEAL3      & LOGN3      & TCSA3      \\ \hline
BRSR6      & MULT3      & STBP3      & LIGT3      & ENEV3      & DMMO3      \\ \hline
TRIS3      & GOAU4      & ECOR3      & ENAT3      & QUAL3      & VLID3      \\ \hline
PRIO3      & LINX3      & MYPK3      & EVEN3      & BPAN4      & BEEF3      \\ \hline
\end{tabular}
\Fonte{Elaborada pelo autor (2020).}
\end{table}

\section{Simulações Realizadas}

As simulações foram feitas para cada ação individualmente, e para cada carteira formada na Seção \ref{sec:organizacao}. Para cada ação ou carteira, foram realizadas 1000 simulações de movimento browniano geométrico usando a Equação \ref{eq:sol_mbg}, simulando os preços no ano de 2019, que teve 247 dias em que o mercado de ações esteve aberto.

\section{Comparação entre preços reais e simulados}

Foram utilizados dois métodos para a comparação entre os preços reais e simulados, o primeiro deles foi o coeficiente de correlação, que produz um valor entre -1 e 1, onde -1 é uma correlação negativa perfeita, 0 é nenhuma correlação e 1 é uma correlação positiva perfeita \cite[~p.29-30]{reddy2016simulating}. A fórmula para o cálculo do coeficiente de correlação ($r$) é mostrada na Equação \ref{eq:corr}, onde $x$ e $y$ são as variáveis que se quer comparar (preços reais e simulados), e $n$ é o número de observações.
\begin{equation}
    r = \frac{n(\sum\limits_{i=1}^{n} xy) - (\sum\limits_{i=1}^{n} x)(\sum\limits_{i=1}^{n} y)}{\sqrt{[n(\sum\limits_{i=1}^{n} x^2) - (\sum\limits_{i=1}^{n} x)^2 ][n(\sum\limits_{i=1}^{n} y^2) - (\sum\limits_{i=1}^{n} y)^2]}}
    \label{eq:corr}
\end{equation}

O segundo método utilizado foi o erro percentual absoluto médio (MAPE), segundo \citeonline[~p. 59]{lawrence2009fundamentals} o MAPE serve como uma avaliação amplamente utilizada em métodos de previsão. A fórmula dele está na Equação \ref{eq:mape}, onde $A_t$ é o preço real da ação, $F_t$ é o preço simulado e $n$ é o número de simulações.
\begin{equation}
    MAPE = \frac{\sum\limits^n_{t=1}\left|\frac{A_t - F_t}{F_t}\right|}{n}
    \label{eq:mape}
\end{equation}

\citeonline{abidin2014forecasting} oferecem uma escala de intervalo para avaliar o MAPE. Essa escala está representada na Tabela \ref{tab:esc_map}.

\begin{table}[!h]
\centering
\small
\captionsetup{width = 6.5cm}
\caption{Escala de intervalo para avaliação do MAPE}
\label{tab:esc_map}
\begin{tabular}{|c|c|}
\hline
\rowcolor[HTML]{C0C0C0} 
MAPE          & Precisão de Previsão \\ \hline
Até 10\%      & Alta precisão        \\ \hline
11\% a 20\%   & Boa previsão         \\ \hline
21\% a 50\%   & Previsão razoável    \\ \hline
Mais que 51\% & Previsão imprecisa   \\ \hline
\end{tabular}
\Fonte{\citeonline{abidin2014forecasting}.}
\end{table}


Os dois métodos de comparação foram analisados para simulações no período de uma semana, duas semanas, um mês, seis meses e um ano.

%% file: 2-textuais/6-resultados.tex
\chapter{Resultados e Discussão}
\label{chap:resultados}

\section{Ações Individuais}

A Tabela \ref{tab:corr_ind} apresenta as médias das correlações para as simulações das ações individualmente com os preços reais para cada período selecionado.

\begin{table}[!h]
\centering
\captionsetup{width=11.7cm}
\caption{Médias dos coeficientes de correlações para simulações individuais de ações}
\label{tab:corr_ind}
\small

\hspace{10cm}(continua)

\begin{tabular}{|c|c|c|c|c|c|}
\hline
\rowcolor[HTML]{C0C0C0} 
Ação            & \multicolumn{1}{c|}{\cellcolor[HTML]{C0C0C0}\textbf{1 semana}} & \multicolumn{1}{c|}{\cellcolor[HTML]{C0C0C0}\textbf{2 semanas}} & \multicolumn{1}{c|}{\cellcolor[HTML]{C0C0C0}\textbf{1 mês}} & \multicolumn{1}{c|}{\cellcolor[HTML]{C0C0C0}\textbf{6 meses}} & \multicolumn{1}{c|}{\cellcolor[HTML]{C0C0C0}\textbf{1 ano}} \\ \hline
\textbf{ABCB4}  & 0,008654                                                       & 0,101539                                                        & 0,161682                                                    & 0,020343                                                      & 0,150864                                                    \\ \hline
\textbf{TIET11} & -0,00277                                                       & -0,01895                                                        & 0,024035                                                    & -0,03713                                                      & -0,07145                                                    \\ \hline
\textbf{ALSO3}  & 0,018659                                                       & 0,03735                                                         & 0,090288                                                    & 0,054687                                                      & 0,205158                                                    \\ \hline
\textbf{ALUP11} & -0,08207                                                       & 0,049868                                                        & 0,071682                                                    & 0,230163                                                      & 0,283936                                                    \\ \hline
\textbf{ANIM3}  & 0,007782                                                       & -0,0036                                                         & 0,00815                                                     & 0,005623                                                      & 0,022794                                                    \\ \hline
\textbf{ARZZ3}  & -0,01595                                                       & -0,04595                                                        & -0,01038                                                    & -0,21358                                                      & 0,279321                                                    \\ \hline
\textbf{BPAN4}  & -0,00613                                                       & -0,0025                                                         & 0,016945                                                    & -0,02052                                                      & -0,09409                                                    \\ \hline
\textbf{BRSR6}  & -0,03925                                                       & 0,109404                                                        & 0,056284                                                    & 0,029668                                                      & -0,25283                                                    \\ \hline
\textbf{BRML3}  & -0,02828                                                       & -0,06131                                                        & 0,051825                                                    & -0,03272                                                      & 0,199629                                                    \\ \hline
\textbf{BRPR3}  & -0,01876                                                       & -0,03733                                                        & -0,03076                                                    & -0,04928                                                      & -0,0682                                                     \\ \hline
\textbf{BRAP4}  & 0,057152                                                       & 0,101372                                                        & -0,06189                                                    & 0,148925                                                      & 0,328043                                                    \\ \hline
\textbf{CESP6}  & 0,001466                                                       & -0,03088                                                        & 0,024965                                                    & 0,081445                                                      & 0,16051                                                     \\ \hline
\textbf{HGTX3}  & -0,01504                                                       & -0,06441                                                        & 0,030689                                                    & -0,04821                                                      & 0,15529                                                     \\ \hline
\textbf{COGN3}  & -0,03223                                                       & -0,01874                                                        & -0,00362                                                    & 0,026202                                                      & -0,03212                                                    \\ \hline
\textbf{CSMG3}  & -0,01199                                                       & 0,007091                                                        & -0,1366                                                     & 0,189389                                                      & 0,412051                                                    \\ \hline
\textbf{RLOG3}  & -0,01713                                                       & 0,049484                                                        & 0,095753                                                    & 0,095511                                                      & 0,329327                                                    \\ \hline
\textbf{CVCB3}  & -0,06778                                                       & -0,11966                                                        & 0,17371                                                     & -0,39796                                                      & -0,48685                                                    \\ \hline
\textbf{CYRE3}  & -0,00114                                                       & 0,018848                                                        & 0,089568                                                    & 0,107625                                                      & 0,299601                                                    \\ \hline
\textbf{DIRR3}  & -0,00355                                                       & 0,068429                                                        & 0,096514                                                    & 0,190993                                                      & 0,286314                                                    \\ \hline
\textbf{DMMO3}  & 0,042509                                                       & 0,105358                                                        & 0,190593                                                    & 0,381209                                                      & 0,299059                                                    \\ \hline
\textbf{DTEX3}  & -0,00934                                                       & 0,013531                                                        & 0,065172                                                    & -0,1209                                                       & 0,161752                                                    \\ \hline
\textbf{ECOR3}  & 0,016021                                                       & 0,081486                                                        & 0,077633                                                    & -0,06056                                                      & 0,240445                                                    \\ \hline
\textbf{EMBR3}  & 0,016257                                                       & -0,01294                                                        & 0,060503                                                    & 0,109184                                                      & 0,147331                                                    \\ \hline
\textbf{ENAT3}  & 0,015454                                                       & -0,01614                                                        & 0,000583                                                    & 0,040881                                                      & 0,066057                                                    \\ \hline
\textbf{ENBR3}  & 0,017443                                                       & 0,057351                                                        & 0,063767                                                    & 0,11467                                                       & 0,191727                                                    \\ \hline
\textbf{ENEV3}  & 0,012342                                                       & -0,00857                                                        & -0,00418                                                    & -0,04075                                                      & -0,05509                                                    \\ \hline
\textbf{EVEN3}  & 0,008616                                                       & 0,005193                                                        & -0,00081                                                    & 0,030115                                                      & 0,086309                                                    \\ \hline
\textbf{EZTC3}  & -0,0415                                                        & -0,01709                                                        & 0,080231                                                    & 0,13187                                                       & 0,386084                                                    \\ \hline
\textbf{FLRY3}  & 0,088561                                                       & 0,085364                                                        & 0,143073                                                    & -0,01026                                                      & 0,4945                                                      \\ \hline
\textbf{GFSA3}  & 0,01595                                                        & 0,057004                                                        & 0,063259                                                    & 0,14395                                                       & 0,142456                                                    \\ \hline
\textbf{GOAU4}  & 0,072696                                                       & 0,002773                                                        & 0,054742                                                    & -0,1072                                                       & 0,043826                                                    \\ \hline
\textbf{GOLL4}  & -0,02561                                                       & -0,00144                                                        & -0,01805                                                    & 0,085167                                                      & 0,223769                                                    \\ \hline
\textbf{GRND3}  & 0,007169                                                       & 0,082445                                                        & 0,111981                                                    & -0,18793                                                      & 0,241728                                                    \\ \hline
\textbf{GUAR3}  & -0,0658                                                        & -0,00967                                                        & -0,02701                                                    & -0,41011                                                      & 0,319611                                                    \\ \hline
\textbf{HBOR3}  & 0,029301                                                       & -0,01964                                                        & -0,06057                                                    & 0,100708                                                      & -0,10098                                                    \\ \hline
\textbf{IGTA3}  & -0,06809                                                       & -0,08895                                                        & 0,069322                                                    & 0,000718                                                      & 0,362464                                                    \\ \hline
\textbf{MEAL3}  & 0,021409                                                       & -0,01533                                                        & -0,03909                                                    & 0,05963                                                       & 0,073278                                                    \\

\end{tabular}
\end{table}

\begin{table}[!h]
\captionsetup{width=11.4cm}
\centering
\small
\hspace{9.6cm}(conclusão)

\begin{tabular}{|c|c|c|c|c|c|}
\hline
\rowcolor[HTML]{C0C0C0} 
Ação            & \multicolumn{1}{c|}{\cellcolor[HTML]{C0C0C0}\textbf{1 semana}} & \multicolumn{1}{c|}{\cellcolor[HTML]{C0C0C0}\textbf{2 semanas}} & \multicolumn{1}{c|}{\cellcolor[HTML]{C0C0C0}\textbf{1 mês}} & \multicolumn{1}{c|}{\cellcolor[HTML]{C0C0C0}\textbf{6 meses}} & \multicolumn{1}{c|}{\cellcolor[HTML]{C0C0C0}\textbf{1 ano}} \\ \hline

\textbf{MYPK3}  & -0,00573                                                       & -0,03686                                                        & -0,04942                                                    & -0,12345                                                      & -0,12982                                                    \\ \hline
\textbf{JHSF3}  & 0,021938                                                       & -0,01548                                                        & 0,007069                                                    & 0,038497                                                      & 0,061186                                                    \\ \hline
\textbf{JSLG3}  & 0,028052                                                       & 0,016545                                                        & -0,04761                                                    & -0,01474                                                      & -0,01289                                                    \\ \hline
\textbf{LIGT3}  & -0,01005 & -0,00399 & 0,015843 & 0,02613  & 0,023918 \\ \hline
\textbf{LINX3}  & -0,05717 & -0,08391 & -0,10439 & 0,124011 & 0,068908 \\ \hline
\textbf{LCAM3}  & -0,11454 & -0,18629 & 0,19647  & 0,573686 & 0,774921 \\ \hline
\textbf{LOGN3}  & 0,005535 & -0,01191 & 0,006302 & -0,0459  & -0,13477 \\ \hline
\textbf{AMAR3}  & -0,00632 & 0,005556 & 0,010618 & -0,03148 & -0,09995 \\ \hline
\textbf{LPSB3}  & -0,01823 & -0,00531 & -0,01347 & -0,14431 & 0,204341 \\ \hline
\textbf{MDIA3}  & 0,050303 & 0,087082 & 0,026938 & -0,20758 & -0,28369 \\ \hline
\textbf{POMO4}  & -0,0105  & -0,01852 & -0,09325 & -0,1541  & -0,09526 \\ \hline
\textbf{MRFG3}  & 0,027532 & 0,017539 & -0,03558 & -0,12099 & -0,13272 \\ \hline
\textbf{LEVE3}  & 0,030932 & 0,007111 & 0,011794 & 0,002021 & 0,013073 \\ \hline
\textbf{MILS3}  & -0,00822 & -0,00106 & 0,00508  & 0,013447 & -0,00062 \\ \hline
\textbf{BEEF3}  & -0,08475 & -0,10494 & -0,1808  & -0,35734 & -0,48876 \\ \hline
\textbf{MRVE3}  & -0,00246 & 0,05614  & 0,083826 & 0,204185 & 0,257913 \\ \hline
\textbf{MULT3}  & -0,04223 & -0,05229 & 0,096332 & 0,056239 & 0,324829 \\ \hline
\textbf{ODPV3}  & -0,02443 & 0,037143 & 0,043043 & 0,107612 & 0,048914 \\ \hline
\textbf{PRIO3}  & 0,045823 & 0,038239 & 0,078428 & 0,159874 & 0,264239 \\ \hline
\textbf{POSI3}  & 0,00897  & 0,005804 & -0,00328 & 0,01194  & -0,02004 \\ \hline
\textbf{QUAL3}  & 0,019193 & -0,01132 & -0,00278 & -0,03543 & -0,07623 \\ \hline
\textbf{RAPT4}  & 0,001824 & -0,00923 & 0,079874 & -0,17523 & 0,264589 \\ \hline
\textbf{SAPR4}  & 0,035799 & 0,069615 & 0,12332  & 0,170788 & 0,468344 \\ \hline
\textbf{STBP3}  & -0,02552 & 0,011441 & 0,01442  & -0,04618 & 0,170859 \\ \hline
\textbf{SMTO3}  & -0,01467 & 0,010901 & 0,042876 & 0,053229 & 0,061823 \\ \hline
\textbf{SEER3}  & 0,015256 & 0,020865 & 0,130148 & 0,118089 & 0,121837 \\ \hline
\textbf{SQIA3}  & -0,02588 & -0,03772 & 0,101281 & 0,268417 & 0,371002 \\ \hline
\textbf{SLCE3}  & -0,06187 & 0,019099 & 0,086921 & -0,25416 & -0,13175 \\ \hline
\textbf{SMLS3}  & 0,002333 & 0,003374 & -0,01143 & 0,005723 & -0,01032 \\ \hline
\textbf{TAEE11} & 0,067337 & 0,05197  & 0,112373 & 0,212627 & 0,388771 \\ \hline
\textbf{TASA4}  & -0,01581 & 0,010986 & -0,00607 & -0,03084 & 0,019806 \\ \hline
\textbf{TCSA3}  & -0,05557 & -0,01354 & 0,085329 & 0,112605 & -0,02539 \\ \hline
\textbf{TGMA3}  & 0,011571 & -0,06524 & 0,104693 & 0,001946 & 0,498978 \\ \hline
\textbf{TEND3}  & 0,111902 & 0,084402 & 0,167719 & 0,368552 & 0,478175 \\ \hline
\textbf{TRIS3}  & 0,136238 & 0,07772  & -0,07531 & 0,349626 & 0,455078 \\ \hline
\textbf{TUPY3}  & -0,00236 & -0,01471 & 0,00758  & -0,00284 & 0,017903 \\ \hline
\textbf{USIM5}  & 0,045209 & 0,021627 & -0,01547 & -0,18045 & -0,1562  \\ \hline
\textbf{VLID3}  & 0,039551 & -0,04791 & -0,1175  & 0,239445 & 0,296091 \\ \hline
\textbf{VULC3}  & -0,00771 & 0,00901  & 0,023455 & -0,05079 & 0,058071 \\ \hline
\textbf{WIZS3}  & 0,019807 & -0,03152 & -0,04369 & -0,09945 & -0,21456 \\ \hline
\textbf{YDUQ3}  & -0,01902 & 0,0118   & 0,05758  & 0,015248 & 0,155584 \\ \hline
\textbf{Média} & 0,000219 & 0,004654  & 0,032042 & 0,023080 & 0,119074 \\ \hline
\end{tabular}
\Fonte{Elaborada pelo autor (2020).}
\end{table}

A Tabela \ref{tab:mape_ind} apresenta o resultado do MAPE, novamente para as simulações das ações individualmente para cada período selecionado.

\begin{table}[!h]
\centering
\small
\captionsetup{width=11.5cm}
\caption{MAPE para ações simuladas individualmente}
\label{tab:mape_ind}
\hspace{10cm}(continua)

\begin{tabular}{|c|c|c|c|c|c|}
\hline
\rowcolor[HTML]{C0C0C0} 
Ação            & \textbf{1 semana} & \textbf{2 semanas} & \textbf{1 mês} & \textbf{6 meses} & \textbf{1 ano} \\ \hline
\textbf{ABCB4}  & 0,055074          & 0,07827            & 0,12562        & 0,157145         & 0,186356       \\ \hline
\textbf{TIET11} & 0,035792          & 0,055261           & 0,094132       & 0,157132         & 0,309047       \\ \hline
\textbf{ALSO3}  & 0,031604          & 0,059058           & 0,094721       & 0,14355          & 0,296411       \\ \hline
\textbf{ALUP11} & 0,023585          & 0,035354           & 0,06963        & 0,209535         & 0,2709         \\ \hline
\textbf{ANIM3}  & 0,107589          & 0,115603           & 0,141805       & 0,210157         & 0,330519       \\ \hline
\textbf{ARZZ3}  & 0,038926          & 0,051744           & 0,075271       & 0,166087         & 0,221031       \\ \hline
\textbf{BPAN4}  & 0,039384          & 0,057231           & 0,099096       & 1,101525         & 2,743757       \\ \hline
\textbf{BRSR6}  & 0,041387          & 0,058146           & 0,078006       & 0,168317         & 0,243293       \\ \hline
\textbf{BRML3}  & 0,034716          & 0,041809           & 0,067215       & 0,134445         & 0,202715       \\ \hline
\textbf{BRPR3}  & 0,028045          & 0,041408           & 0,06719        & 0,162137         & 0,370399       \\ \hline
\textbf{BRAP4}  & 0,047868          & 0,062841           & 0,099528       & 0,205619         & 0,283937       \\ \hline
\textbf{CESP6}  & 0,034502          & 0,043151           & 0,061622       & 0,196108         & 0,304434       \\ \hline
\textbf{HGTX3}  & 0,058573          & 0,074327           & 0,10475        & 0,235777         & 0,345182       \\ \hline
\textbf{COGN3}  & 0,054188          & 0,088264           & 0,174531       & 0,274647         & 0,406283       \\ \hline
\textbf{CSMG3}  & 0,049381          & 0,054784           & 0,069521       & 0,153853         & 0,212917       \\ \hline
\textbf{RLOG3}  & 0,040405          & 0,055198           & 0,118971       & 0,219725         & 0,347317       \\ \hline
\textbf{CVCB3}  & 0,036155          & 0,049227           & 0,064987       & 0,206813         & 0,313012       \\ \hline
\textbf{CYRE3}  & 0,040577          & 0,055739           & 0,086348       & 0,158555         & 0,363173       \\ \hline
\textbf{DIRR3}  & 0,090109          & 0,1057             & 0,144313       & 0,23646          & 0,42455        \\ \hline
\textbf{DMMO3}  & 0,126414          & 0,168449           & 0,24483        & 0,645996         & 2,561031       \\ \hline
\textbf{DTEX3}  & 0,038608          & 0,051439           & 0,076008       & 0,174802         & 0,227663       \\ \hline
\textbf{ECOR3}  & 0,055702          & 0,083603           & 0,134491       & 0,182565         & 0,295031       \\ \hline
\textbf{EMBR3}  & 0,035359          & 0,044282           & 0,074268       & 0,149828         & 0,2052         \\ \hline
\textbf{ENAT3}  & 0,190427          & 0,179179           & 0,212002       & 0,617849         & 0,756185       \\ \hline
\textbf{ENBR3}  & 0,042166          & 0,053942           & 0,110648       & 0,251559         & 0,314423       \\ \hline
\textbf{ENEV3}  & 0,06318           & 0,084476           & 0,126173       & 0,316653         & 0,75561        \\ \hline
\textbf{EVEN3}  & 0,04498           & 0,059086           & 0,097098       & 0,190975         & 0,568403       \\ \hline
\textbf{EZTC3}  & 0,037881          & 0,045791           & 0,072426       & 0,145085         & 0,37609        \\ \hline
\textbf{FLRY3}  & 0,055412          & 0,07275            & 0,091357       & 0,129044         & 0,183861       \\ \hline
\textbf{GFSA3}  & 0,041998          & 0,055826           & 0,089417       & 0,428367         & 0,470878       \\ \hline
\textbf{GOAU4}  & 0,066008          & 0,084038           & 0,107247       & 0,225792         & 0,318724       \\ \hline
\textbf{GOLL4}  & 0,071918          & 0,0958             & 0,131459       & 0,302958         & 0,477893       \\ \hline
\textbf{GRND3}  & 0,02483           & 0,03839            & 0,062289       & 0,122663         & 0,173735       \\ \hline
\textbf{GUAR3}  & 0,03835           & 0,046196           & 0,065786       & 0,222533         & 0,26991        \\ \hline
\textbf{HBOR3}  & 0,063173          & 0,086185           & 0,13894        & 0,218588         & 0,862981       \\ \hline
\textbf{IGTA3}  & 0,029183          & 0,038632           & 0,05091        & 0,121484         & 0,153797       \\ \hline
\textbf{MEAL3}  & 0,036464          & 0,046714           & 0,067086       & 0,151961         & 0,233709       \\ \hline
\textbf{MYPK3}  & 0,038776          & 0,051379           & 0,077298       & 0,185822         & 0,257146       \\ \hline
\textbf{JHSF3}  & 0,055077          & 0,067279           & 0,092424       & 0,346261         & 1,075657       \\ \hline
\textbf{JSLG3}  & 0,083464          & 0,120968           & 0,205888       & 0,590452         & 1,325502       \\ \hline
\textbf{LIGT3}  & 0,057603          & 0,093632           & 0,142653       & 0,270143         & 0,34758        \\ \hline
\textbf{LINX3}  & 0,049501          & 0,062861           & 0,085565       & 0,15214          & 0,196829       \\ \hline
\textbf{LCAM3}  & 0,044845          & 0,061456           & 0,076431       & 0,171607         & 0,236604       \\ \hline
\textbf{LOGN3}  & 0,079359          & 0,090516           & 0,126373       & 0,361362         & 1,323929       \\ \hline
\textbf{AMAR3}  & 0,090847 & 0,108189 & 0,136458 & 0,372294 & 0,783193 \\ 
\end{tabular}
\end{table}

\begin{table}[!h]
\centering
\small
\captionsetup{width=11.5cm}
\hspace{9.2cm}(conclusão)

\begin{tabular}{|c|c|c|c|c|c|}
\hline

\textbf{LPSB3}  & 0,054853 & 0,072617 & 0,094421 & 0,262187 & 0,338588 \\ \hline
\textbf{MDIA3}  & 0,073172 & 0,085825 & 0,092097 & 0,143129 & 0,213129 \\ \hline
\textbf{POMO4}  & 0,05011  & 0,055143 & 0,07004  & 0,169584 & 0,234526 \\ \hline
\textbf{MRFG3}  & 0,044121 & 0,052614 & 0,087137 & 0,244281 & 0,62924  \\ \hline
\textbf{LEVE3}  & 0,04121  & 0,055986 & 0,082999 & 0,132867 & 0,189605 \\ \hline
\textbf{MILS3}  & 0,05397  & 0,077738 & 0,208925 & 0,322974 & 0,660096 \\ \hline
\textbf{BEEF3}  & 0,06378  & 0,084937 & 0,139792 & 0,590528 & 1,246437 \\ \hline
\textbf{MRVE3}  & 0,033634 & 0,066069 & 0,11495  & 0,209807 & 0,360769 \\ \hline
\textbf{MULT3}  & 0,029956 & 0,034235 & 0,048366 & 0,111906 & 0,15478  \\ \hline
\textbf{ODPV3}  & 0,03595  & 0,049699 & 0,10613  & 0,194091 & 0,22879  \\ \hline
\textbf{PRIO3}  & 0,078038 & 0,106035 & 0,145168 & 0,546718 & 0,688567 \\ \hline
\textbf{POSI3}  & 0,075205 & 0,082152 & 0,106165 & 0,227687 & 0,916045 \\ \hline
\textbf{QUAL3}  & 0,057603 & 0,086898 & 0,163936 & 0,413086 & 1,100659 \\ \hline
\textbf{RAPT4}  & 0,047582 & 0,060223 & 0,08508  & 0,192731 & 0,26356  \\ \hline
\textbf{SAPR4}  & 0,12926  & 0,163782 & 0,224137 & 0,357677 & 0,467723 \\ \hline
\textbf{STBP3}  & 0,062752 & 0,08577  & 0,131974 & 0,277122 & 0,548933 \\ \hline
\textbf{SMTO3}  & 0,026604 & 0,034982 & 0,067118 & 0,11035  & 0,154562 \\ \hline
\textbf{SEER3}  & 0,056876 & 0,091973 & 0,216281 & 0,388375 & 0,542144 \\ \hline
\textbf{SQIA3}  & 0,035037 & 0,046461 & 0,067002 & 0,293302 & 0,741082 \\ \hline
\textbf{SLCE3}  & 0,041446 & 0,05462  & 0,076968 & 0,15317  & 0,231359 \\ \hline
\textbf{SMLS3}  & 0,046974 & 0,060775 & 0,086348 & 0,235452 & 0,301884 \\ \hline
\textbf{TAEE11} & 0,045666 & 0,056297 & 0,070075 & 0,112033 & 0,16117  \\ \hline
\textbf{TASA4}  & 0,750423 & 0,894669 & 0,59932  & 0,389304 & 0,50173  \\ \hline
\textbf{TCSA3}  & 0,086131 & 0,102113 & 0,095676 & 0,165454 & 0,272468 \\ \hline
\textbf{TGMA3}  & 0,042989 & 0,053169 & 0,074488 & 0,179166 & 0,237006 \\ \hline
\textbf{TEND3}  & 0,044139 & 0,054914 & 0,083295 & 0,145365 & 0,247087 \\ \hline
\textbf{TRIS3}  & 0,065492 & 0,084365 & 0,091112 & 0,224869 & 0,699406 \\ \hline
\textbf{TUPY3}  & 0,031593 & 0,041222 & 0,057575 & 0,128699 & 0,175827 \\ \hline
\textbf{USIM5}  & 0,086982 & 0,110236 & 0,132923 & 0,253575 & 0,345967 \\ \hline
\textbf{VLID3}  & 0,038429 & 0,05017  & 0,106361 & 0,174042 & 0,222689 \\ \hline
\textbf{VULC3}  & 0,071853 & 0,126876 & 0,172307 & 0,303665 & 0,442408 \\ \hline
\textbf{WIZS3}  & 0,041911 & 0,085374 & 0,12717  & 0,444069 & 0,874587 \\ \hline
\textbf{YDUQ3}  & 0,058379 & 0,114176 & 0,200501 & 0,253174 & 0,428327 \\ \hline
\textbf{Média} & 0,063481          & 0,082773           & 0,114854       & 0,255087         & 0,483948       \\ \hline
\end{tabular}
\Fonte{Elaborada pelo autor (2020).}
\end{table}

\begin{figure}[h!] 
   	    \captionsetup{width=15cm}
		\Caption{\label{fig:simulacoes} Exemplos de simulações realizadas}
		\ifcefig{}{
			\includegraphics[width=15cm]{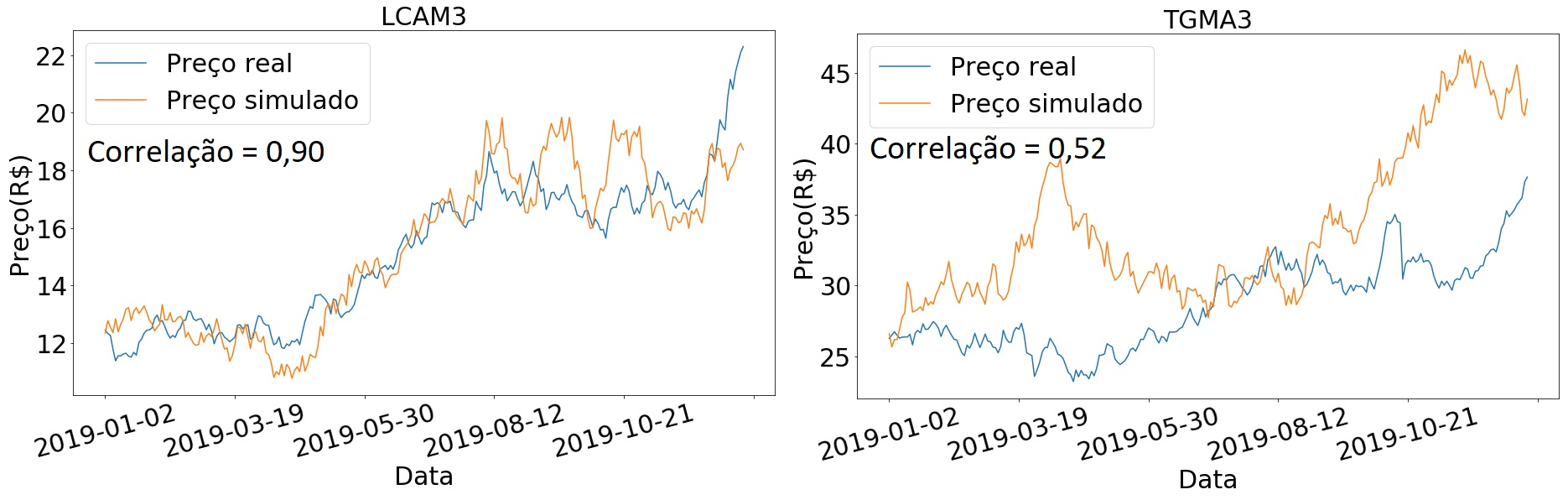}
		}{
			\Fonte{Elaborada pelo autor (2020).}
		}	
	\end{figure}

A Figura \ref{fig:simulacoes} mostra exemplos de simulações de movimento browniano geométrico realizadas para duas ações, comparadas com os preços reais delas, sendo que para cada ação foram realizadas 1000 simulações deste tipo.

Para estas simulações, no geral, os coeficientes de correlação não mostram bons resultados. Em média, a correlação fica mais forte para simulações de maiores períodos, mas ainda assim, sem correlações fortes.

Analisando o MAPE, que segundo \citeonline[~p.36]{reddy2016simulating}, são os resultados de maior valor que o coeficiente de correlação para movimento browniano geométrico, segundo a escala de \citeonline{abidin2014forecasting}, em média as simulações foram altamente precisas para uma e duas semanas, para um mês foram boas previsões, e para seis meses e um ano foram previsões razoáveis.

\section{Carteiras ordenadas por retornos}

A Tabela \ref{tab:corr_return} apresenta a média das correlações para as simulações das carteiras ordenadas por retorno esperado, que estão detalhadas na Tabela \ref{tab:cart_retorno}.

\begin{table}[!h]
\centering
\small
\captionsetup{width = 15.2cm}
\caption{Média dos coeficientes de correlação para carteiras ordenadas por retorno}
\label{tab:corr_return}
\begin{tabular}{c|c|c|c|c|c|c|}
\cline{2-6}
\multicolumn{1}{l|}{}                     &
\multicolumn{1}{c|}{\cellcolor[HTML]{C0C0C0}\textbf{\begin{tabular}[c]{@{}c@{}}Retorno\\ Anual\end{tabular}}} &
\multicolumn{1}{c|}{\cellcolor[HTML]{C0C0C0}\textbf{1 semana}} & \multicolumn{1}{c|}{\cellcolor[HTML]{C0C0C0}\textbf{2 semanas}} & \multicolumn{1}{c|}{\cellcolor[HTML]{C0C0C0}\textbf{1 mês}} & \multicolumn{1}{c|}{\cellcolor[HTML]{C0C0C0}\textbf{6 meses}} & \multicolumn{1}{c|}{\cellcolor[HTML]{C0C0C0}\textbf{1 ano}} \\ \hline
\multicolumn{1}{|c|}{\textbf{\begin{tabular}[c]{@{}c@{}}carteira 1\\ (maiores retornos)\end{tabular}}} & 69\% & 0,007704                                                       & 0,1731                                                          & 0,315672                                                    & 0,471653                                                      & 0,771296                                                    \\ \hline
\multicolumn{1}{|c|}{\textbf{carteira 2}} & 42\% & -0,00608                                                       & 0,163421                                                        & 0,021401                                                    & -0,03847                                                      & 0,713684                                                    \\ \hline
\multicolumn{1}{|c|}{\textbf{carteira 3}} & 28\% & -0,05816                                                       & 0,084554                                                        & 0,177263                                                    & 0,104113                                                      & 0,471752                                                    \\ \hline
\multicolumn{1}{|c|}{\textbf{carteira 4}} & 20\% & -0,01651                                                       & 0,059439                                                        & 0,141538                                                    & 0,278535                                                      & 0,414327                                                    \\ \hline
\multicolumn{1}{|c|}{\textbf{carteira 5}} & 10\% & -0,00537                                                       & 0,059178                                                        & 0,079065                                                    & 0,138313                                                      & 0,214799                                                    \\ \hline
\multicolumn{1}{|c|}{\textbf{\begin{tabular}[c]{@{}c@{}}carteira 6\\ (menores retornos)\end{tabular}}} & -7\% & 0,029056                                                       & -0,07144                                                        & -0,07496                                                    & -0,15901                                                      & -0,26016                                                    \\ \hline
\end{tabular}
\Fonte{Elaborada pelo autor (2020).}
\end{table}

A Tabela \ref{tab:mape_return} mostra como ficou o MAPE para as mesmas simulações de cada carteira nos períodos de tempo escolhidos.

\begin{table}[!h]
\centering
\small
\captionsetup{width = 15.3cm}
\caption{MAPE para carteiras ordenadas por retorno}
\label{tab:mape_return}
\begin{tabular}{c|c|c|c|c|c|c|}
\cline{2-6}
\multicolumn{1}{l|}{}                     &
\multicolumn{1}{c|}{\cellcolor[HTML]{C0C0C0}\textbf{\begin{tabular}[c]{@{}c@{}}Retorno\\ Anual\end{tabular}}} &
\multicolumn{1}{c|}{\cellcolor[HTML]{C0C0C0}\textbf{1 semana}} & \multicolumn{1}{c|}{\cellcolor[HTML]{C0C0C0}\textbf{2 semanas}} & \multicolumn{1}{c|}{\cellcolor[HTML]{C0C0C0}\textbf{1 mês}} & \multicolumn{1}{c|}{\cellcolor[HTML]{C0C0C0}\textbf{6 meses}} & \multicolumn{1}{c|}{\cellcolor[HTML]{C0C0C0}\textbf{1 ano}} \\ \hline
\multicolumn{1}{|c|}{\textbf{\begin{tabular}[c]{@{}c@{}}carteira 1\\ (maiores retornos)\end{tabular}}} & 69\% & 0,036328                                                       & 0,043452                                                        & 0,056674                                                    & 0,12213                                                       & 0,179603                                                    \\ \hline
\multicolumn{1}{|c|}{\textbf{carteira 2}} & 42\% & 0,071972                                                       & 0,085288                                                        & 0,07443                                                     & 0,099478                                                      & 0,141729                                                    \\ \hline
\multicolumn{1}{|c|}{\textbf{carteira 3}} & 28\% & 0,040842                                                       & 0,048816                                                        & 0,07202                                                     & 0,099338                                                      & 0,139845                                                    \\ \hline
\multicolumn{1}{|c|}{\textbf{carteira 4}} & 20\% & 0,029566                                                       & 0,037095                                                        & 0,064005                                                    & 0,13079                                                       & 0,203834                                                    \\ \hline
\multicolumn{1}{|c|}{\textbf{carteira 5}} & 10\% & 0,043204                                                       & 0,053921                                                        & 0,085964                                                    & 0,138117                                                      & 0,406951                                                    \\ \hline
\multicolumn{1}{|c|}{\textbf{\begin{tabular}[c]{@{}c@{}}carteira 6\\ (menores retornos)\end{tabular}}} & -7\% & 0,030017                                                       & 0,044102                                                        & 0,067255                                                    & 0,211474                                                      & 0,536942                                                    \\ \hline
\end{tabular}
\Fonte{Elaborada pelo autor (2020).}
\end{table}

\begin{figure}[h!] 
   	    \captionsetup{width=15cm}
		\Caption{\label{fig:simulacoes_carteira1_return} Aplicação do movimento browniano geométrico para carteiras 1 e 6 (ordenadas por retornos)}
		\ifcefig{}{
			\includegraphics[width=15cm]{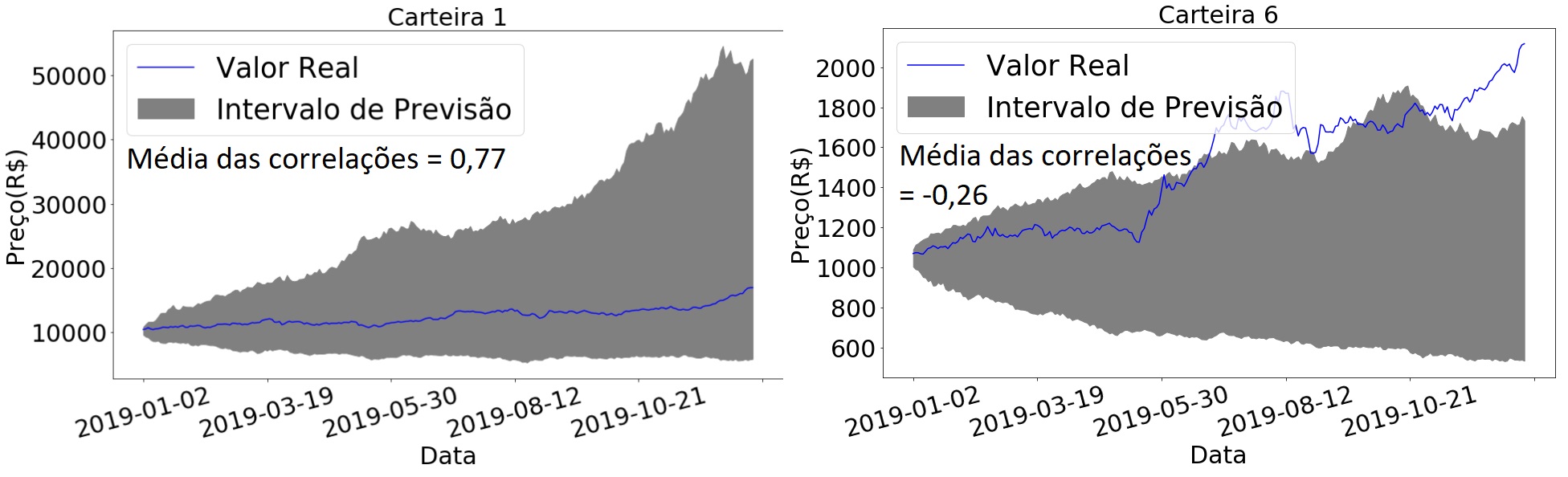}
		}{
			\Fonte{Elaborada pelo autor (2020).}
		}	
\end{figure}

A Figura \ref{fig:simulacoes_carteira1_return} mostra os intervalos de previsões para as carteira 1 e 6, que são respectivamente as de melhor e pior resultado no período determinado de acordo com o MAPE.

\begin{figure}[h!] 
   	    \captionsetup{width=15cm}
		\Caption{\label{fig:mape_return} Variação do MAPE ao longo do tempo para carteiras ordenadas por retorno.}
		\ifcefig{}{
			\includegraphics[width=15cm]{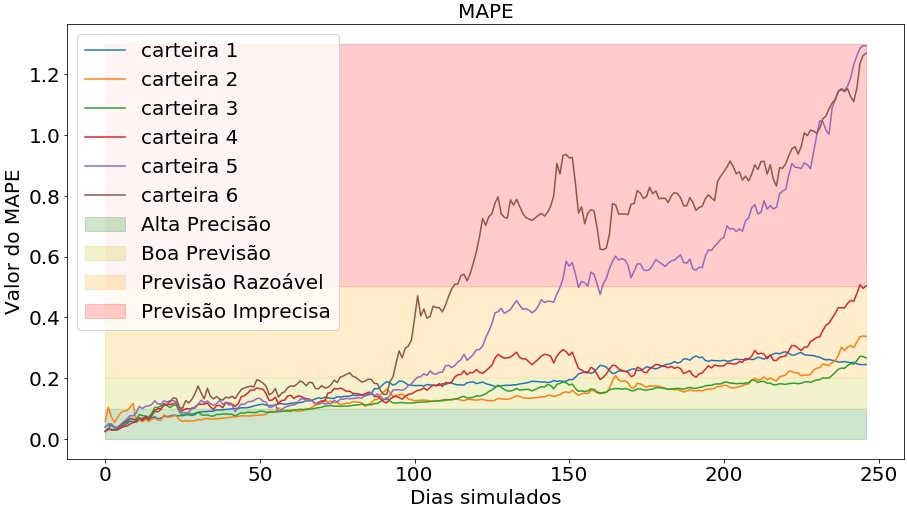}
		}{
			\Fonte{Elaborada pelo autor (2020).}
		}	
\end{figure}

Nesse caso, os coeficientes de correlação foram bem maiores principalmente nas primeiras carteiras para longos períodos, já para curto prazos e para carteiras de baixo retorno a correlação é quase nula e as vezes até negativa. Isso pode se dar por o movimento browniano geométrico com o passar do tempo ir compensando flutuações aleatórias negativas, já que no geral, os preços de ações tendem a subir com o tempo \cite[~p. 33]{reddy2016simulating}.

O MAPE também mostrou resultados melhores em comparação com as simulações de ações individualmente. Seguindo a escala de \citeonline{abidin2014forecasting}, as simulações de uma semana até um mês em todos os casos, e nas de seis meses para as carteiras 2 e 3 foram altamente precisas, nas de 6 meses para as carteiras 1, 4 e 5 foram boas previsões e razoável para a 6, nas simulações de 1 ano, foram boas para as carteiras 1, 2 e 3, razoável nas carteiras 4 e 5, e imprecisas para a 6. A Figura \ref{fig:mape_return} mostra como o MAPE variou ao longo do tempo para as simulações de cada carteira, fica fácil notar como as carteiras de maiores retornos tiveram resultados melhores, e que para curtos períodos de tempo o MAPE se manteve baixo em todas.

Isso mostra que simulações para carteiras de alto retorno esperado são mais precisas. Considerando que o MAPE está sendo o parâmetro principal de avaliação, as simulações foram melhores nas carteiras de alto retorno em períodos de até seis meses.

\section{Carteiras ordenadas por risco}

A Tabela \ref{tab:mcorr_vol} mostra as médias dos coeficientes de correlação para as carteiras ordenadas por risco, que foram detalhadas na Tabela \ref{tab:cart_risco} e na Tabela \ref{tab:mmape_vol} estão os valores do MAPE para as simulações.

\begin{table}[!h]
\centering
\small
\captionsetup{width = 14.5cm}
\caption{Média dos coeficientes de correlação para carteiras ordenadas por risco}
\label{tab:mcorr_vol}
\begin{tabular}{c|c|c|c|c|c|c|}
\cline{2-6}
                                          &
                                  \cellcolor[HTML]{C0C0C0}\textbf{\begin{tabular}[c]{@{}c@{}}Risco\\ Anual\end{tabular}} &        \cellcolor[HTML]{C0C0C0}\textbf{1 semana} & \cellcolor[HTML]{C0C0C0}\textbf{2 semanas} & \cellcolor[HTML]{C0C0C0}\textbf{1 mês} & \cellcolor[HTML]{C0C0C0}\textbf{6 meses} & \cellcolor[HTML]{C0C0C0}\textbf{1 ano} \\ \hline
\multicolumn{1}{|c|}{\textbf{\begin{tabular}[c]{@{}c@{}}carteira 1\\ (maiores riscos)\end{tabular}}} & 36\% & 0,047237                                  & 0,145909                                   & 0,107035                               & 0,171145                                 & 0,602854                               \\ \hline
\multicolumn{1}{|c|}{\textbf{carteira 2}} & 27\% & 0,022307                                  & 0,076255                                   & 0,102442                               & 0,260057                                 & 0,42456                                \\ \hline
\multicolumn{1}{|c|}{\textbf{carteira 3}} & 23\% & 0,024329                                  & 0,082379                                   & 0,202267                               & 0,411145                                 & 0,578064                               \\ \hline
\multicolumn{1}{|c|}{\textbf{carteira 4}} & 22\% & -0,10612                                  & -0,06600                                   & 0,217767                               & 0,42246                                  & 0,687747                               \\ \hline
\multicolumn{1}{|c|}{\textbf{carteira 5}} & 19\% & -0,08567                                  & 0,018578                                   & 0,248532                               & -0,11362                                 & 0,613941                               \\ \hline
\multicolumn{1}{|c|}{\textbf{\begin{tabular}[c]{@{}c@{}}carteira 6\\ (menores riscos)\end{tabular}}} & 16\% & 0,033036                                  & 0,101719                                   & 0,163674                               & 0,276202                                 & 0,55112                                \\ \hline
\end{tabular}
\Fonte{Elaborada pelo autor (2020).}
\end{table}

\begin{table}[!h]
\centering
\small
\captionsetup{width = 14.5cm}
\caption{MAPE para carteiras ordenadas por risco}
\label{tab:mmape_vol}
\begin{tabular}{c|c|c|c|c|c|c|}
\cline{2-6}
                                          &
 \cellcolor[HTML]{C0C0C0}\textbf{\begin{tabular}[c]{@{}c@{}}Risco\\ Anual\end{tabular}} &         \cellcolor[HTML]{C0C0C0}\textbf{1 semana} & \cellcolor[HTML]{C0C0C0}\textbf{2 semanas} & \cellcolor[HTML]{C0C0C0}\textbf{1 mês} & \cellcolor[HTML]{C0C0C0}\textbf{6 meses} & \cellcolor[HTML]{C0C0C0}\textbf{1 ano} \\ \hline
\multicolumn{1}{|c|}{\textbf{\begin{tabular}[c]{@{}c@{}}carteira 1\\ (maiores riscos)\end{tabular}}}& 36\%  & 0,086591                                  & 0,106456                                   & 0,112916                               & 0,155108                                 & 0,201853                               \\ \hline
\multicolumn{1}{|c|}{\textbf{carteira 2}}& 27\% & 0,035099                                  & 0,046088                                   & 0,065676                               & 0,113360                                  & 0,188864                               \\ \hline
\multicolumn{1}{|c|}{\textbf{carteira 3}}& 23\% &  0,032039                                  & 0,040036                                   & 0,054298                               & 0,123337                                 & 0,291293                               \\ \hline
\multicolumn{1}{|c|}{\textbf{carteira 4}}& 22\% &  0,028574                                  & 0,030137                                   & 0,041559                               & 0,088433                                 & 0,129947                               \\ \hline
\multicolumn{1}{|c|}{\textbf{carteira 5}}& 19\% &  0,021811                                  & 0,025615                                   & 0,037230                                & 0,086436                                 & 0,111154                               \\ \hline
\multicolumn{1}{|c|}{\textbf{\begin{tabular}[c]{@{}c@{}}carteira 6\\ (menores riscos)\end{tabular}}}& 16\%  & 0,024279                                  & 0,035298                                   & 0,062621                               & 0,083606                                 & 0,124045                               \\ \hline
\end{tabular}
\Fonte{Elaborada pelo autor (2020).}
\end{table}

\begin{figure}[h!] 
   	    \captionsetup{width=15cm}
		\Caption{\label{fig:simulacoes_carteira1_vol} Aplicação do movimento browniano geométrico para carteiras  5 e 3 (ordenadas por riscos)}
		\ifcefig{}{
			\includegraphics[width=15cm]{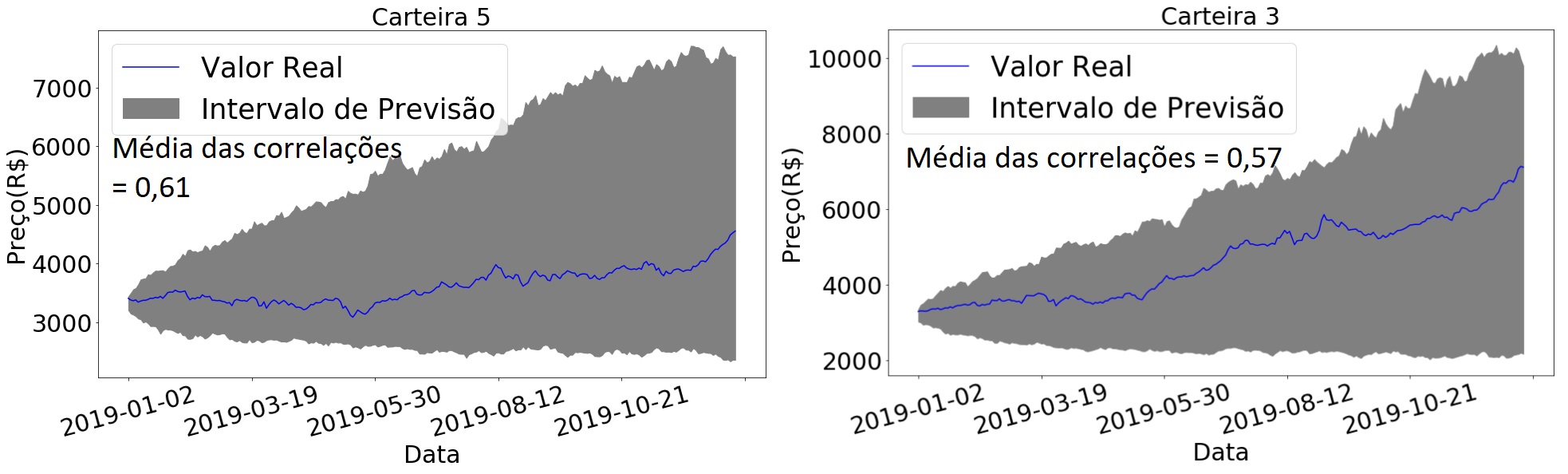}
		}{
			\Fonte{Elaborada pelo autor (2020).}
		}	
\end{figure}

A Figura \ref{fig:simulacoes_carteira1_vol} mostra os intervalos de previsão para as carteiras 5 e 3 que foram respectivamente as de melhor e pior resultado com base no MAPE.

\begin{figure}[h!] 
   	    \captionsetup{width=15cm}
		\Caption{\label{fig:mape_risk} Variação do MAPE ao longo do tempo para carteiras ordenadas por risco.}
		\ifcefig{}{
			\includegraphics[width=15cm]{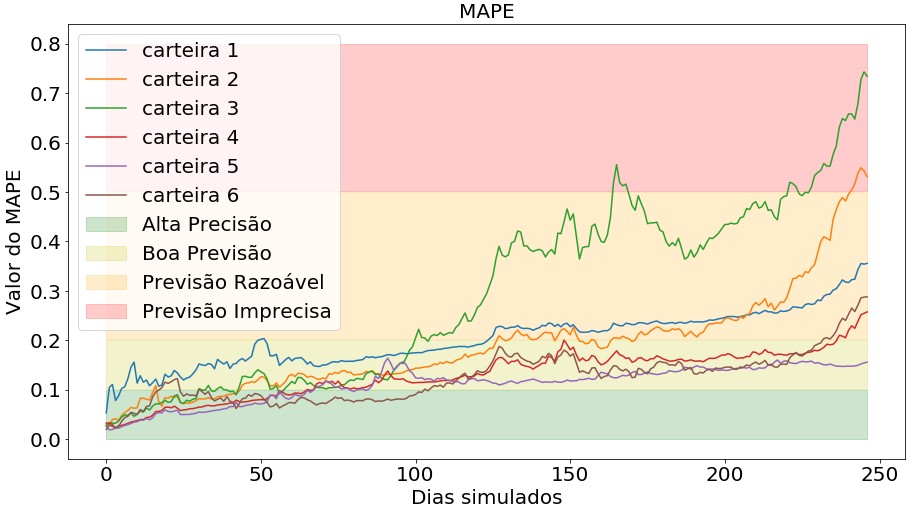}
		}{
			\Fonte{Elaborada pelo autor (2020).}
		}	
\end{figure}
	
Analisando os coeficientes de correlação, dessa vez não foi notada diferença significativa dos coeficientes das diferentes carteiras, então, diferente do caso das carteiras ordenadas por retorno, as volatilidades dessas carteiras não parecem afetar diretamente nas correlações, mas como antes, a correlação fica mais forte para simulações de tempos mais longos.

Já o MAPE foi afetado pelos riscos. É notável que as carteiras de menores riscos tiveram melhores resultados, as carteiras 3, 4 e 5 tiveram resultados altamente precisos para simulações de até seis meses, e boa previsão para um ano, as carteiras 2 e 3 tiveram simulações altamente precisas para o tempo de até um mês, e boas previsões para seis meses, para um ano, a carteira 2 teve boa previsão, e a carteira 3 teve previsão razoável, e a carteira 1, que foi composta pelas ações mais voláteis, só teve resultado altamente preciso para uma semana, entre duas semanas e seis meses teve boa previsão, e previsão razoável para um ano. A Figura \ref{fig:mape_risk} mostra a variação do MAPE ao longo dos dias simulados, ela torna mais fácil notar as carteiras com melhores e piores simulações, além de novamente ser possível perceber que os resultados são melhores para curtos períodos.

De acordo com o MAPE, as simulações tiveram um melhor desempenho para carteiras menos voláteis, e como nos casos anteriores, para períodos menores.

\section{Carteiras otimizadas ordenadas pelo IS}

A Tabela \ref{tab:mcorr_sharpe} mostra as médias dos coeficientes de correlação para as carteiras ordenadas pelo \gls{IS} e otimizadas utilizando as teorias de \citeonline{doi:10.1111/j.1540-6261.1952.tb01525.x} e \citeonline{sharpe1963simplified}.

\begin{table}[!h]
\centering
\small
\captionsetup{width = 15.2cm}
\caption{Média dos coeficientes de correlação para carteiras ordenadas por IS}
\label{tab:mcorr_sharpe}
\begin{tabular}{c|c|c|c|c|c|c|}
\cline{2-6}
\multicolumn{1}{l|}{}                     &
\cellcolor[HTML]{C0C0C0}\textbf{\begin{tabular}[c]{@{}c@{}}Índice\\ de Sharpe\end{tabular}} & 
\multicolumn{1}{c|}{\cellcolor[HTML]{C0C0C0}\textbf{1 semana}} & \multicolumn{1}{c|}{\cellcolor[HTML]{C0C0C0}\textbf{2 semanas}} & \multicolumn{1}{c|}{\cellcolor[HTML]{C0C0C0}\textbf{1 mês}} & \multicolumn{1}{c|}{\cellcolor[HTML]{C0C0C0}\textbf{6 meses}} & \multicolumn{1}{c|}{\cellcolor[HTML]{C0C0C0}\textbf{1 ano}} \\ \hline
\multicolumn{1}{|c|}{\textbf{\begin{tabular}[c]{@{}c@{}}carteira 1\\ (maiores Índices \\ de Sharpe)\end{tabular}}} & 2,80 & -0,03888                                                       & 0,09203                                                         & 0,339554                                                    & 0,709637                                                      & 0,838319                                                    \\ \hline
\multicolumn{1}{|c|}{\textbf{carteira 2}}& 1,50 & -0,08971                                                       & -0,10613                                                        & 0,177518                                                    & 0,496156                                                      & 0,643859                                                    \\ \hline
\multicolumn{1}{|c|}{\textbf{carteira 3}}& 1,30 & -0,05596                                                       & 0,086511                                                        & 0,191161                                                    & 0,092304                                                      & 0,584119                                                    \\ \hline
\multicolumn{1}{|c|}{\textbf{carteira 4}}& 0,96 & 0,003985                                                       & 0,111891                                                        & 0,132509                                                    & 0,25133                                                       & 0,466629                                                    \\ \hline
\multicolumn{1}{|c|}{\textbf{carteira 5}}& 0,28 & -0,01436                                                       & 0,029958                                                        & 0,029018                                                    & 0,059099                                                      & 0,105898                                                    \\ \hline
\multicolumn{1}{|c|}{\textbf{\begin{tabular}[c]{@{}c@{}}carteira 6\\ (menores Índices \\ de Sharpe)\end{tabular}}}& -0,20 & 0,003601                                                       & -0,02097                                                        & -0,03447                                                    & -0,07298                                                      & -0,12866                                                    \\ \hline
\end{tabular}
\Fonte{Elaborada pelo autor (2020).}
\end{table}

Na Tabela \ref{tab:mmape_sharpe} são exibidos os valores para o MAPE das carteiras.

\begin{table}[!h]
\centering
\small
\captionsetup{width = 15.2cm}
\caption{MAPE para carteiras ordenadas por IS}
\label{tab:mmape_sharpe}
\begin{tabular}{c|c|c|c|c|c|c|}
\cline{2-6}
                                          &
                                        \cellcolor[HTML]{C0C0C0}\textbf{\begin{tabular}[c]{@{}c@{}}Índice\\ de Sharpe\end{tabular}} &  \cellcolor[HTML]{C0C0C0}\textbf{1 semana} & \cellcolor[HTML]{C0C0C0}\textbf{2 semanas} & \cellcolor[HTML]{C0C0C0}\textbf{1 mês} & \cellcolor[HTML]{C0C0C0}\textbf{6 meses} & \cellcolor[HTML]{C0C0C0}\textbf{1 ano} \\ \hline
\multicolumn{1}{|c|}{\textbf{\begin{tabular}[c]{@{}c@{}}carteira 1\\ (maiores Índices \\ de Sharpe)\end{tabular}}} & 2,80 & 0,040444                                  & 0,043613                                   & 0,059183                               & 0,110034                                 & 0,153657                               \\ \hline
\multicolumn{1}{|c|}{\textbf{carteira 2}} & 1,50 & 0,024163                                  & 0,029471                                   & 0,042701                               & 0,096837                                 & 0,19784                                \\ \hline
\multicolumn{1}{|c|}{\textbf{carteira 3}} & 1,30 & 0,030421                                  & 0,036708                                   & 0,055897                               & 0,094675                                 & 0,146321                               \\ \hline
\multicolumn{1}{|c|}{\textbf{carteira 4}} & 0,96 & 0,079783                                  & 0,108132                                   & 0,130438                               & 0,139891                                 & 0,186187                               \\ \hline
\multicolumn{1}{|c|}{\textbf{carteira 5}} & 0,28 & 0,043408                                  & 0,052257                                   & 0,085423                               & 0,364667                                 & 0,990568                               \\ \hline
\multicolumn{1}{|c|}{\textbf{\begin{tabular}[c]{@{}c@{}}carteira 6\\ (menores Índices \\ de Sharpe)\end{tabular}}} & -0,20 & 0,038402                                  & 0,055098                                   & 0,084792                               & 0,172935                                 & 0,472557                               \\ \hline
\end{tabular}
\Fonte{Elaborada pelo autor (2020).}
\end{table}

\begin{figure}[h!] 
   	    \captionsetup{width=15cm}
		\Caption{\label{fig:simulacoes_carteira1_sharpe} Aplicação do movimento browniano geométrico para as carteiras 3 e 5 (ordenadas por IS)}
		\ifcefig{}{
			\includegraphics[width=15cm]{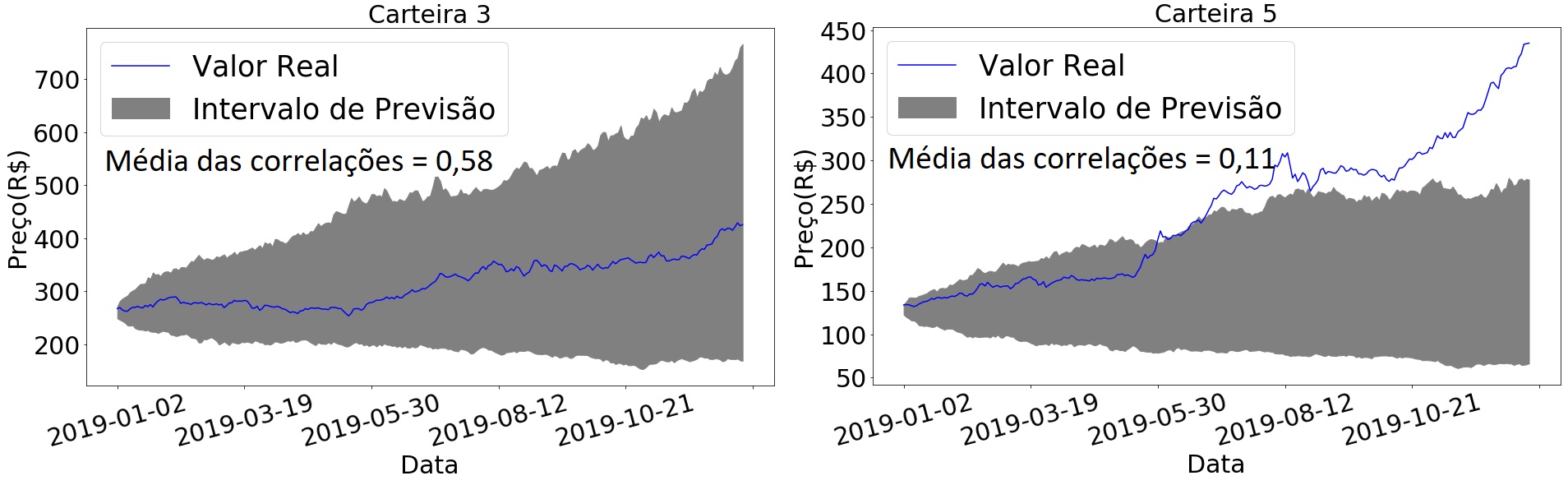}
		}{
			\Fonte{Elaborada pelo autor (2020).}
		}	
\end{figure}

A Figura \ref{fig:simulacoes_carteira1_sharpe} mostra os intevalos de previsões para as carteiras 3 e 5 que foram respectivamente as de melhor e pior resultado com base no MAPE.

\begin{figure}[h!] 
   	    \captionsetup{width=15cm}
		\Caption{\label{fig:mape_is} Variação do MAPE ao longo do tempo para carteiras ordenadas IS.}
		\ifcefig{}{
			\includegraphics[width=15cm]{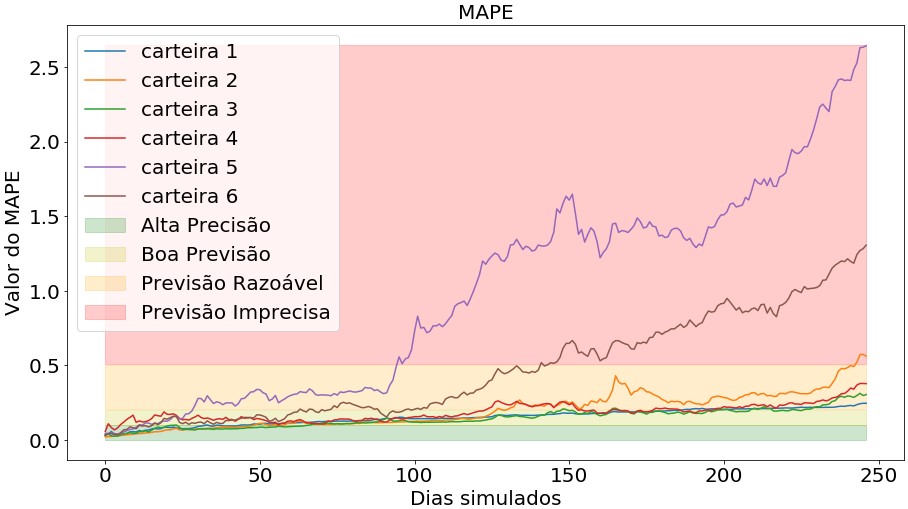}
		}{
			\Fonte{Elaborada pelo autor (2020).}
		}	
\end{figure}

Observando os coeficientes de correlação, é possível notar que a carteira 1 obteve a maior média de coeficientes de todos os testes realizados, e quanto menor é o \gls{IS}, menor são as correlações. Assim como nos casos anteriores, no geral, a correlação aumenta para períodos maiores.

E de acordo com o MAPE, carteiras com maiores \gls{IS} novamente obtiveram melhores resultados, principalmente para curtos períodos. A carteira 1 teve simulação altamente precisa para o período de até um mês, e boas previsões para seis meses a um ano, as carteiras 2 e 3 foram altamente precisas para períodos de até seis meses, e apresentaram boas previsões para um ano, a carteira 4 foi altamente precisa somente para uma semana, e de duas semanas a um ano obteve boas previsões, a carteira 5 teve simulações altamente precisas para o período de até um mês, previsão regular para seis meses, e sem precisão para um ano. Por fim, a carteira 6 teve resultados altamente precisos para até um mês, boa previsão para seis meses, e previsão regular para um ano. A Figura \ref{fig:mape_is} mostra como o MAPE variou ao longo do tempo, sendo possível mais uma vez notar as carteiras de melhores e piores resultados, assim como que para curtos períodos o MAPE se manteve menor.

Os bons resultados dessas carteiras corroboram de acordo com os resultados anteriores, já que foi possível concluir que carteiras com alto retorno e baixo risco tem melhores resultados, e o \gls{IS} é proporcional ao retorno, e inversamente proporcional a volatilidade. Também é interessante notar o quanto as carteiras 5 e 6, que são as de ações com menores \gls{IS}, tiveram resultados bem imprecisos em comparação com as outras, tanto na análise por coeficiente de correlação quando na análise por MAPE.

%% file: 2-textuais/7-conclusao.tex
\chapter{Conclusão}
\label{chap:conclusao}

Com este estudo foi possível explorar o movimento browniano geométrico para simulação de preços de ações listadas no Índice de \emph{Small Caps} da bolsa de valores brasileira em quatro situações diferentes. A primeira situação foi a simulação de ações individuais, foram utilizados dois métodos de verificação dos resultados, o primeiro, que foi o coeficiente de correlação, não mostrou bons resultados, porém, o segundo método, que foi o erro percentual absoluto médio (MAPE), que é o mais significativo para este tipo de simulações, mostrou melhores resultados principalmente para uma e duas semanas de previsão.

A segunda situação analisada foi a simulação de carteiras ordenadas de acordo com o retorno de suas ações. Foram obtidas correlações mais fortes do que na simulação de carteiras individuais. Para as carteiras que contém as ações de maior retorno esperado, mesmo com fraca correlação para períodos pequenos de simulação, essa correlação vai se tornando mais forte a medida em que o tempo aumenta, isso mostra que para períodos longos, os comportamentos das simulações e dos valores reais são semelhantes, não necessariamente com grandes taxas de acerto, mas com precisão razoável sobre a direção em que as carteiras seguirão. A análise por meio do MAPE mostrou resultados mais precisos para curtos períodos, mas diferente da simulação de carteiras individuais, os resultados sugeriram boa precisão também para períodos um pouco maiores como de um e seis meses, principalmente nas carteiras de maior retorno.

Para a terceira situação, foram analisadas a simulação de carteiras ordenadas pelo risco de suas ações. Os coeficientes de correlação não foram alterados significativamente por diferentes volatilidades, mas continuaram maiores para períodos mais longos assim como no segundo caso, ou seja, mostrando novamente comportamento parecido com o dos valores reais. O MAPE se mostrou maior para as carteiras de menores riscos, novamente com melhores precisões para períodos de até seis meses .

O último caso analisado foi a simulação de carteiras ordenadas pelos \gls{IS} de suas ações, mas dessa vez, em vez de ser igualmente balanceada para cada ação, ela foi otimizada com os modelos de \citeonline{doi:10.1111/j.1540-6261.1952.tb01525.x} e \citeonline{sharpe1963simplified}. No período de um ano, a carteira de maior \gls{IS} mostrou o maior coeficiente de correlação desse trabalho, sendo mais de 0,8, uma correlação forte, ou seja, para este período, foi a simulação com comportamento geral mais semelhante ao comportamento do preço real. O MAPE, assim como na segunda e terceira situação, se mostrou melhor para curtos prazos, e para carteiras de maiores \gls{IS}.

No trabalho de \citeonline{reddy2016simulating}, foram simuladas carteiras ordenadas por risco e retorno, mas com ações de índices macro do mercado australiano, no geral eles obtiveram resultados mais precisos, o que pode se dar pelo menor risco das ações desses índices ultilizados por eles.

Os resultados obtidos foram interessantes, pois esse tipo de simulação pode ser utilizado por investidores para melhorar seus rendimentos, além de que não foram encontrados estudos desse tipo utilizando ações listadas como \emph{Small Caps}. Também foi interessante observar como a diversificação do investimento através de diferentes carteiras parece influenciar os resultados. Com perspectivas futuras, poderão ser testadas simulações com outros agrupamentos de carteiras, como o agrupamento por setores dentro dos índices, também poderão ser testadas simulações baseadas em diferentes períodos do histórico de dados.